\documentclass[twocolumn]{aastex63}

\newcommand{\eg}{e.g.,}

\newcommand{\Msun}{$M_{\odot}$}

\newcommand{\kms}{km~s$^{-1}$}
\newcommand{\ergs}{erg s$^{-1}$}

\newcommand{\OI}{O~{\sc i}}

\newcommand{\NaI}{Na~{\sc i}}

\newcommand{\MgII}{Mg~{\sc ii}}

\newcommand{\SiII}{Si~{\sc ii}}

\newcommand{\SII}{S~{\sc ii}}
\newcommand{\CaII}{Ca~{\sc ii}}
\newcommand{\TiII}{Ti~{\sc ii}}

\newcommand{\FeII}{Fe~{\sc ii}}
\newcommand{\FeIII}{Fe~{\sc iii}}
\newcommand{\CoII}{Co~{\sc ii}}

\newcommand{\NiII}{Ni~{\sc ii}}

\newcommand{\Nifs}{$^{56}$Ni}

\newcommand{\sBV}{s$_{BV}$}

\newcommand{\DmB}{$\Delta \mathrm{m}_{15}(B)$}
\newcommand{\DmV}{$\Delta \mathrm{m}_{15}(V)$}

\newcommand{\ved}{$v_{edge}$}
\newcommand{\iBmax}{t$^{i-B}_{max}$}

\newcommand{\Ia}{SNe Ia}
\newcommand{\bg}{1991bg-like SNe}
\newcommand{\snpy}{SNooPy}
\newcommand{\bo}{SN~2015bo}
\newcommand{\cn}{SN~1997cn}
\newcommand{\timeB}{t$(B){_\mathrm{max}}$}

\definecolor{mygray}{gray}{0.9}

\newcommand{\nothing}[1]{#1}

\shorttitle{A Tale of Two Supernovae}
\shortauthors{Hoogendam et al.}

\graphicspath{{./}{figures/}}

\begin{document}

\title{A Tale of Two Type Ia Supernovae: The fast-declining siblings SNe 2015bo and  1997cn}

\author[0000-0003-3953-9532]{W. B. Hoogendam}
\affiliation{Institute for Astronomy, University of Hawaii, 2680 Woodlawn Drive, Honolulu, HI 96822, USA}
\affiliation{Department of Physics and Astronomy, Calvin University, Grand Rapids, MI 49546, USA}

\author[0000-0002-5221-7557]{C. Ashall}
\affiliation{Institute for Astronomy, University of Hawaii, 2680 Woodlawn Drive, Honolulu, HI 96822, USA}

\author[0000-0002-1296-6887]{L. Galbany}
\affiliation{Institute of Space Sciences (ICE, CSIC), Campus UAB, Carrer de Can Magrans, s/n, E-08193 Barcelona, Spain.}
\affiliation{Institut d’Estudis Espacials de Catalunya (IEEC), E-08034 Barcelona, Spain.}

\author[0000-0003-4631-1149]{B. J. Shappee}
\affiliation{Institute for Astronomy, University of Hawaii, 2680 Woodlawn Drive, Honolulu, HI 96822, USA}

\author[0000-0003-4625-6629]{C. R. Burns}
\affiliation{Observatories of the Carnegie Institution for Science, 813 Santa Barbara St., Pasadena, CA 91101, USA}

\author[0000-0002-3900-1452]{J. Lu}
\affiliation{Department of Physics, Florida State University, 77 Chieftan Way, Tallahassee, FL 32306, USA}

\author[0000-0003-2734-0796]{M. M. Phillips}
\affiliation{Carnegie Observatories, Las Campanas Observatory, Casilla 601, La Serena, Chile}

\author[0000-0001-5393-1608]{E. Baron}
\affiliation{8Homer L. Dodge Department of Physics and Astronomy, University of Oklahoma, 440 W. Brooks, Rm 100, Norman, OK 73019-2061,USA}
\affiliation{Hamburger Sternwarte, Gojenbergsweg 112, D-21029 Hamburg, Germany}

\author[0000-0002-3415-322X]{S. Holmbo}
\affiliation{Department of Physics and Astronomy, Aarhus University, Ny Munkegade 120, DK-8000 Aarhus C, Denmark}

\author[0000-0003-1039-2928]{E. Y. Hsiao}
\affiliation{Department of Physics, Florida State University, 77 Chieftan Way, Tallahassee, FL 32306, USA}

\author[0000-0003-2535-3091]{N. Morrell}
\affiliation{Carnegie Observatories, Las Campanas Observatory, Casilla 601, La Serena, Chile}

\author[0000-0002-5571-1833]{M. D. Stritzinger}
\affiliation{Department of Physics and Astronomy, Aarhus University, Ny Munkegade 120, DK-8000 Aarhus C, Denmark}

\author[0000-0002-8102-181X]{N. B. Suntzeff}
\affiliation{George P. and Cynthia Woods Mitchell Institute for Fundamental Physics \& Astronomy, Texas A\&M University, Department of Physics
and Astronomy, 4242 TAMU, College Station, TX 77843, USA}

\author[0000-0002-2387-6801]{F. Taddia}
\affiliation{Department of Physics and Astronomy, Aarhus University, Ny Munkegade 120, DK-8000 Aarhus C, Denmark}

\author[0000-0002-1229-2499]{D. R. Young}
\affiliation{Astrophysics Research Centre, School of Mathematics and Physics, Queen's University Belfast, Belfast BT7 1NN, UK}

\author[0000-0002-3464-0642]{J. D. Lyman}
\affiliation{Department of Physics, University of Warwick, Gibbet Hill Road, Coventry CV4 7AL, UK}

\author[0000-0002-3256-0016]{S. Benetti}
\affiliation{INAF - Osservatorio Astronomico di Padova, Vicolo dell'Osservatorio 5, I-35122 Padova, Italy}

\author[0000-0001-6876-8284]{P. A. Mazzali}
\affiliation{Astrophysics Research Institute, Liverpool John Moores University, IC2, Liverpool Science Park, 146 Brownlow Hill, Liverpool L3 5RF, UK}
\affiliation{Max-Planck-Institut für Astrophysik, Karl-Schwarzschild Str 1, D-85748 Garching, Germany}

\author{M. Delgado Manche\~no}
\affiliation{Departamento de F\'isica Te\'orica y del Cosmos, Universidad de Granada, E-18071 Granada, Spain}
\author{R. Gonz\'alez D\'iaz}
\affiliation{Instituto Nacional de Astro\'isica, \'Optica y Electr\'onica (INAOE), 72840 Tonantzintla, Puebla, Mexico.}
\affiliation{Institute of Space Sciences (ICE, CSIC), Campus UAB, Carrer de Can Magrans, s/n, E-08193 Barcelona, Spain.}
\author{S. Mu\~noz Torres}
\affiliation{Departamento de F\'isica Te\'orica y del Cosmos, Universidad de Granada, E-18071 Granada, Spain}

\begin{abstract}
We present optical and near-infrared photometric and spectroscopic observations of the fast-declining Type Ia Supernova (SN) 2015bo. 
\bo\ is under-luminous ($M_B$~=~\nothing{$-17.67$}~$\pm$~0.15~mag) and has a fast-evolving light curve (\DmB~=~1.91~$\pm$~0.01~mag and \sBV~=~0.48~$\pm$~0.01). 
It has a unique morphology in the observed $V-r$ color curve, where it is bluer than all other SNe in the comparison sample. A \Nifs\ mass of 0.17~$\pm$~0.03~\Msun\ was derived from the peak bolometric luminosity, which is consistent with its location on the luminosity-width relation. Spectroscopically, SN~2015bo is a Cool SN in the Branch classification scheme. The velocity evolution measured from spectral features is consistent with 1991bg-like SNe. SN~2015bo has a SN twin (similar spectra) \emph{and} sibling (same host galaxy), SN~1997cn. Distance \nothing{moduli} of $\mu$~=~34.33~$\pm$~0.01~(stat)~$\pm$~0.11~(sys)~mag and $\mu$~=~34.34~$\pm$~0.04~(stat)~$\pm$~0.12~(sys)~mag are derived for SN~2015bo and SN~1997cn, respectively. These distances are consistent at the 0.06$-\sigma$ level with each other, and they are also consistent with distances derived using surface-brightness fluctuations and redshift-corrected cosmology. This suggests that fast-declining SNe could be accurate distance indicators which should not be excluded from future cosmological analyses. 
\end{abstract}



\section{Introduction}\label{sec:intro}

Type Ia supernovae (hereafter \Ia) have many astrophysical and cosmological applications. They are one of the major sources of metal enrichment \citep{Raiteri96}, used as standard candles \citep[e.g.][]{Phillips93,Phillips99,Burns18}, and they established the acceleration rate of the  universe \citep[e.g.][]{riess98, perl99}. \Ia\ have additional applications: they continue to be used to calculate the Hubble constant \citep[e.g.][]{Riess16, Burns18,khet21}, improve constraints on the equation-of-state for dark energy $w$ parameter \citep[e.g.][]{beto14,scol18}, and investigate the distribution of dark matter in galaxies \citep[e.g.][]{fein13}.

It is apparent that \Ia\ are not an entirely uniform class of objects; rather, there are a growing number of sub-types. These include: the over-luminous 1991T-like \Ia\  \citep[\eg][]{fil92b,phi92}; 2003fg-like \Ia\ which are brighter in the NIR and have long rise times \citep[\eg][]{how06,hic07,hsiao20,Ashall21,Lu21}; 2002cx-like \Ia\ show light curves that are both broad and faint \citep[\eg][]{li03,fol13}; 2006bt-like \Ia\ are with broad, slowly declining light curves seen in high luminosity \Ia, yet they lack a prominent secondary maximum in the $i$ band as seen in low luminosity \Ia\ \citep[\eg][]{fol10}; 2002ic-like \Ia\ that exhibit Balmer emission lines found in Type IIn core-collapse SNe \citep{ ham03}; and lastly, 1991bg-like \Ia\ are sub-luminous \citep[\eg][]{fil92a, leib93,Turatto96}. In addition to these broad categories, there are also several types of  \Ia\ that straddle the boundaries between these sub-types. One of these is the so-called transitional \Ia. Transitional SNe bridge the gap between normal and \bg. They tend to be fast-declining and exhibit other features of \bg, but are not entirely consistent with the \bg\ classification. Examples of transitional objects include SN~1986G, SN~2003hv, SN~2004eo, SN~2007on, SN~2011iv, iPTF13ebh, and SN~2015bp \citep[\eg][]{Phillips87,Leloudas09, Pastorello07,Gall18,Ashall18, Hsiao15, Wyatt21}. For a full review of all sub-types, see \citet{Taubenberger17}.

Roughly 18\% of all \Ia\ are considered \bg\ \citep{Li2011}. It is hotly debated if they are produced by Chandrasekhar (Ch) mass or sub-Ch mass explosions \citep[\eg][]{Hoeflich02,Stritzinger06, Blondin18}. In addition to being sub-luminous ($M_{B}>-18$~mag; \citealt{Taubenberger17}), \bg\ differ in many ways compared to normal \Ia. For example, \bg\ light curves differ photometrically with their faster rise and decline and less luminous peak (hence the classification as sub-luminous). Whereas other supernovae have either a shoulder or a second maximum in the $i$ band, \bg\ have neither \citep{Ashall20}. However, it is in their spectroscopic properties where \bg\ differ the most: they have noticeably stronger \TiII\ and \OI\ features than other types, and also have a strong \SiII\ feature at 5972~\AA\ \citep{Nugent95,Benetti05}.

One might suspect that, due to their astronomical importance, \Ia\ would be well understood; yet the progenitor systems of \Ia, their explosion details, and the underlying physical interpretation of the empirical relationships are not fully known despite decades of research. 
\Ia\ are agreed to come from the explosion of at least one carbon-oxygen (C-O) white dwarf (WD) \citep{hoy60}. However, there are several potential progenitor scenarios that could produce a SN Ia. These are: the single degenerate (SD) scenario which consists of a single C-O WD and a non-degenerate companion star \citep{Whelan73,Livne90,Nomoto97}, the double degenerate (DD) scenario  which consists of two C-O WDs \citep{iben84,Webbink84}, and the triple/quaternary scenario which consists of two C-O WDs in a system with at least one other companion star \citep{thom11,Pejcha13,Shappee13}. 

For each of these scenarios, there are a variety of viable explosion mechanisms and ejecta masses. 
A C-O WD may accrete mass from a companion star until it approaches the Chandrasekhar limit,  where compressional heating in the center of the explosion can start a thermonuclear disruption \citep[\eg][]{Diamond18}. This  can occur in the SD \citep{Whelan73} or DD scenarios \citep{Piersanti03}. 
Alternatively, a sub-Chandrasekhar mass C-O WD may accrete a surface helium layer which can detonate and produce an inward-moving shock wave and total disruption of the WD \citep{Nomoto80,woosley94,Hoeflich:Khokhlov:96}. This is possible in both the SD or DD scenarios. A sub-Chandrasekhar mass explosion may also result from a merger between two C-O WDs (whose combined mass does not exceed the Chandrasekhar limit; \citealt{vanKerkwijk10}). 
Another explosion mechanism in the DD progenitor scenario is where two white dwarfs violently merge as a result of angular momentum loss from gravitational radiation. The heat generated from merging provides the necessary spark for the powder keg \citep{iben84}.
Finally, the head on collision of two WDs, which may be in a multiple system, can possibly produce a \Ia\ where the explosion is triggered by a detonation wave \citep{Rosswog09,Raskin09,thom11,Katz12,Kushnir13,Pejcha13,Dong15,maz18}.

One of the most important empirical relationships that \Ia\ follow is the luminosity-width relationship. This study utilizes the luminosity-width relation in two representations. First, by examining the difference in $B$-band magnitude from maximum light (\timeB) to fifteen days past maximum (the \DmB\ quantity) against the absolute $B$-band magnitude (M$_{B}$) \citep{Phillips93,Phillips99}. Second, by examining the time difference between the time of $B$-band maximum and intrinsic $B-V$ maximum normalized by 30 days (\sBV) versus M$_{B}$ \citep{Burns14}. For both of these relationships, more luminous objects have broader light curves, however, \sBV\ is a better indicator for fast-declining objects such as \bg\ because \DmB\ is not a reliable measure of the decline rate for values greater than 1.7~mag \citep{Phillips12,Burns14,Gall18}. The physical understanding of the luminosity-width relation lies in the amount of  $^{56}$Ni synthesized in the explosion \citep{Nugent95}. More luminous objects produce more \Nifs\ \citep{Stritzinger06}, which provides more line opacity and increases the diffusion time scales in the ejecta. This causes broader light curves \citep[\eg][]{Mazzali07}. In the case of the least luminous objects, the \Nifs\ mass will be lower and line opacity small which causes quickly evolving light curves.

\Ia\ serve as excellent tools to calculate extra-galactic distances \citep{Phillips93,Hamuy96,Phillips99} after correcting for the luminosity-width relationship \citep[c.f.][]{Hamuy96}. However, analyses of distances for \Ia\ in galaxies with low redshifts ($z\le 0.1$) have additional variance due to peculiar velocities and host galaxy bulk flows relative to cosmological expansion. This can be avoided by comparing \Ia\ that explode in the same galaxy (and thus have the same bulk flow and peculiar velocity) which are called ``siblings" \citep{Brown15}. Calculating distances using siblings can also help mitigate systematic uncertainties stemming from the host-galaxy properties \citep{Kelly10,Sullivan10,Lampeitl10}. \nothing{Additionally, \citet{Scolnic20} used siblings to investigate correlations between SNe Ia light curves and their host galaxies.} Recent studies on \Ia\ siblings include four SNe in NGC~1316 (Fornax~A) \citep{Stritzinger10}, SN~2007on and SN~2011iv in NGC 1404 \citep{Gall18}, SN2013aa and SN~2017cbv in NGC~5643 \citep{Burns20}, and \nothing{\citet{Biswas21} investigated AT~2019lcj and SN~2020aewj}.

\citet{Fakhouri15} introduced the term ``twin" supernovae. Twins are SNe that show analogous spectral features that imply parallel progenitor scenarios and explosion mechanisms. While \citet{Fakhouri15} were able to significantly reduce absolute magnitude differences using twins, another study by \citet{Foley20} found a significant difference in absolute magnitude for the two twins SN~2011by and SN~2011fe. Additionally, SN~2007on and SN~2011iv were also found to be inconsistent with each other despite being in the same galaxy \citep{Gall18}. \cn\ \citep{Turatto98} and \bo\ are spectrally similar fast-declining SNe in NGC 5490, and the existence of these two supernovae presents a rare opportunity to use the power of supernova siblings and twins to determine distances that do not suffer from a slew of systematic uncertainties.

In this paper, we present a detailed multi-wavelength analysis of the fast-declining SN~2015bo that is structured as follows: §\ref{sec:data} presents the data used and a brief analysis of the host galaxy. Light and color curves are derived and compared to other \Ia\ in §\ref{sec:phot}. Spectra are presented in §\ref{sec:spec} and compared to several samples of other sub-types of SNe and sub-luminous SNe. Doppler velocity measurements and pseudo-equivalent width calculations are also presented in §\ref{sec:spec}. In §\ref{sec:dist}, we determine the distance to \bo\ and compare it to the distance derived from the photometry of SN~1997cn: another fast-declining SN which exploded in the same host galaxy. Final discussion and concluding remarks are offered in §\ref{sec:conclusion}. Throughout this work, phases are reported relative to rest-frame $B$-band maximum and given values are corrected for foreground galactic extinction.



\section{Data} \label{sec:data}
SN~2015bo (a.k.a. PSNJ14095513+1731556) was discovered at $\alpha=14^h09^m55^s.130$, $\delta = +17^\circ31'55''.60$ by \citet{howerton16} on 2015-02-14 at 10:37:06 UT with an apparent $V$-band magnitude of $m_V=$~18.4~mag. 

There were two last non-detections; one from the Catalina Real-Time Transient Survey (CRTS; \citet{Drake09} and the other from the All-Sky Automated Survey for SuperNovae (ASAS-SN; \citealt{Shappee14,Kochanek17}) on  2015-01-28 at 10:22:17 UT and 2015-02-08 at 13:47:20.40 UT, respectively. Their respective magnitude limits were $m_V=20.4$ and $m_g=17.76$. SN~2015bo was classified as a \bg\ Ia by the \textit{Public ESO Spectroscopic Survey of Transient Objects} (PESSTO; \citealt{Smartt15}) based on a spectrum taken on 2015-02-15 at 08:29:43 UT \citep{yaron17}. 
It exploded in the type E galaxy NGC 5490 at a redshift of $z=0.0161$\footnote{\citet{vdBosch15} from ned.ipac.caltech.edu.}.
The foreground Galactic extinction towards SN~2015bo is $E(B-V)_{mw}=0.023$~mag \citep{Schlafly11}. \nothing{Photometric data for SN~1997cn was taken from \citet{jha06} and the spectrum in Fig. \ref{fig:bg-spec-comp} is from \citet{tur98}. Spectroscopic data for \cn\ are also available from \citet{Matheson01}, \citet{Blondin12}, and \citet{Silverman12}.} Table \ref{tab:summary} contains properties of both \cn\ and \bo, as well as their host galaxy. \nothing{In the following subsections, we detail the data for \bo.} 

\begin{figure}
\plotone{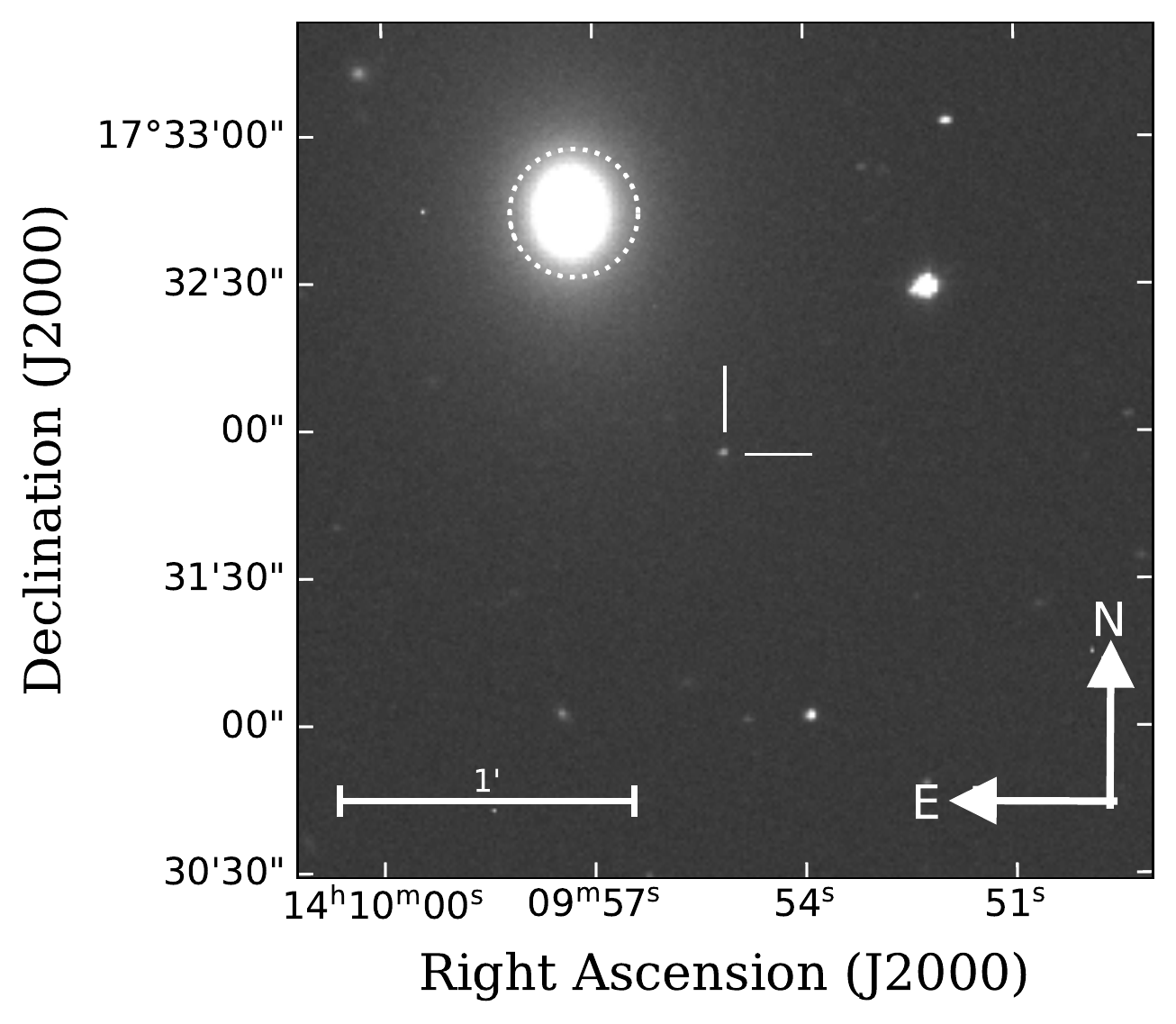}
\caption{Finder chart for \bo\ constructed from an $i$-band image obtained with the Swope telescope on 2015-04-19 at 08:56:55 UT. \label{fig:pic}}
\end{figure}

Data for this work comes from two projects: the \textit{Carnegie Supernova Project II} (CSP-II; \citealt{Phillips19}) and PESSTO \citep{Smartt15}. 
CSP-II ran from 2011-2015 with the aim of obtaining high-precision follow-up observations of SNe, mainly using telescopes at Las Campanas Observatory. PESSTO's objective was to obtain high quality spectral observations of transient sources using the European Southern Observatory's New Technology Telescope (NTT) in La Silla.

\begin{deluxetable}{ccc}
\tablenum{1}
\tablecaption{Properties of SN 1997cn, SN~2015bo and their host galaxy, NGC~5490.  \label{tab:summary}}
\tablewidth{0pt}
\tablehead{
\colhead{Parameters} & \colhead{Value} & \colhead{Ref.}
}
\startdata
\emph{SN~2015bo}: \\
RA (J2000) & $14^h09^m55^s.13$ & (1) \\
           & 212.47971 (deg) & (1) \\
Dec (J2000) & $+17^\circ31'55''.60$& (1) \\
            & 17.53211 (deg) & (1) \\
t$(B){_\mathrm{max}}$ (JD) & 2457078.3 $\pm$ 0.1~(d)  & This work\\
$M_{B}$\tablenotemark{a} & \nothing{$-17.67$} $\pm$ \nothing{0.15} (mag) & This work\\
 $E(B-V)_{MW}$ & 0.023 (mag) & (2) \\
 $E(B-V)_{Host}$ & 0.00 $\pm$ 0.00 (mag) & This work\\
\sBV\tablenotemark{b}  & 0.48 $\pm$ 0.01 & This work \\
\DmB\tablenotemark{c} & 1.91 $\pm$ 0.01 (mag) & This work \\
\hline
\emph{SN~1997cn}: \\
RA (J2000) & $14^h09^m57^s.76$ & (7) \\
           & 212.49067 (deg) & (7) \\
Dec (J2000) & $+17^\circ32'32''.32$& (7) \\
            & +17.54231 (deg) & (7) \\
t$(B){_\mathrm{max}}$ (JD) & 2450588.3 $\pm$ 0.1~(d)  & This work\\
$M_{B}$\tablenotemark{e} & \nothing{$-17.60$} (mag) & This work\\
 $E(B-V)_{MW}$ & 0.023 (mag) & (2) \\
 $E(B-V)_{Host}$ & 0.03 (mag) & (7)\\
\sBV\tablenotemark{b}  & 0.35 $\pm$ 0.06 & This work \\
\DmB\tablenotemark{c} & 1.90 $\pm$ 0.05 (mag) & (7) \\
\hline
\emph{NGC 5490}: \\
Type & Elliptical & (3) \\
RA (J2000) & $14^h09^m57^s.295$ & (4)\\
           & $212.488728$ (deg) & (4)\\
Dec (J2000) & $+17^\circ32^m43^s.98$ & (4)\\
            & 17.545551 (deg) & (4)\\
Velocity, \emph{v} & 12~\kms & (5) \\
$z_{helio}$ & 0.0161 $\pm$ 0.0001 & (6) \\
$\mu$\tablenotemark{d} & 34.21 $\pm$ 0.15 (mag) & (6) \\
\enddata
\textbf{References:}(1) \citet{howerton16}; (2) \citet{Schlafly11}; (3) \citet{third-ref-catalogue-of-galaxies}; (4) \citet{SDSS-DR6}; (5) \citet{Simien97}; (6) \citet{Kowalski08}; (7) \citet{jha06} \\
\tablenotemark{a}{Corrected only for Galactic extinction using $R_V=3.1$}\\
\tablenotemark{b}{Obtained using a Gaussian process fit in SNooPY.}\\
\tablenotemark{c}{Obtained using direct Gaussian process interpolation to the $B$-band. }\\
\tablenotemark{d}{Corrected for Virgo, GA, and Shapely local peculiar motion and calculated using $H_{0}$=73\,km\,s$^{-1}$\,Mpc$^{-1}$, $\Omega_{m}$=0.27, and $\Omega_{\Lambda}$=0.73, which is used throughout this work.}\\
\tablenotemark{e}{This value is from light curve models fit to the data rather than directly from the data.}
\end{deluxetable}

\subsection{Photometric Observations}\label{sec:data-phot}
Photometric observations were acquired in the $uBVgriJHY$ filters from $-$6.9~d to +105.8~d relative to rest-frame $B$-band maximum light. Optical photometry was obtained from the Swope telescope, and the near infrared (NIR) photometry came from the du Pont telescope. Both telescopes are located at the Las Campanas Observatory. All photometric measurements are presented in the natural system and reduced using the methods described in \citet{Krisciunas17} and \citet{Phillips19}. Logs of the optical and NIR observations are presented in Tables \ref{tab:appendix-phot-data-optical} and \ref{tab:appendix-phot-data-NIR}. Host-galaxy template subtraction was not required for \bo\ as it is located far away from the center of the host \emph{at $\sim$20~kpc}.

\subsection{Spectroscopic Observations}\label{sec:data-spec}
Sixteen optical and one NIR spectra of SN~2015bo were obtained from $-$7.8~d to +31.8~d relative to rest-frame $B$-band maximum light. Spectra were taken with a variety of instruments including SPRAT mounted on the Liverpool Telescope, ALFOSC on the Nordic Optical Telescope, EFOSC on the NTT, and WFCCD on the du Pont telescope. The optical spectra were reduced using the standard \textsc{iraf}\footnote{The Image Reduction and Analysis Facility (IRAF) is distributed by the National Optical Astronomy Observatory, which is operated by the Association of Universities for Research in Astronomy, Inc., under cooperative agreement with the National Science Foundation.} packages and methods described in \citet{Hamuy06}, \citet{Folatelli13}, and \citet{Smartt15}. The NIR spectrum was obtained with the Folded-port InfraRed Echellette (FIRE) on the Baade telescope at the Magellan observatory and reduced using the methods described in \citet{Hsiao19}. Table \ref{tab:spec} presents a summary of the spectral data. 

\subsection{Host Galaxy}\label{sec:host-galaxy}

Integral-field spectroscopy of NGC 5490 was obtained on 2021-01-27 08:04:06, with the Multi-Unit Spectroscopic Explorer (MUSE; \citealt{Bacon10}). MUSE is mounted on the Unit 4 telescope at the ESO Very Large Telescope at the the Cerro Paranal Observatory. These host observations were obtained as part of the All-weather MUSE Supernova Integral-field Nearby Galaxies (AMUSING; \citealt{Galbany16,LopezCoba20}) survey. AMUSING is an ongoing project that analyzes the host galaxies of SNe to study their environments.

\begin{figure}
\includegraphics[width=\columnwidth]{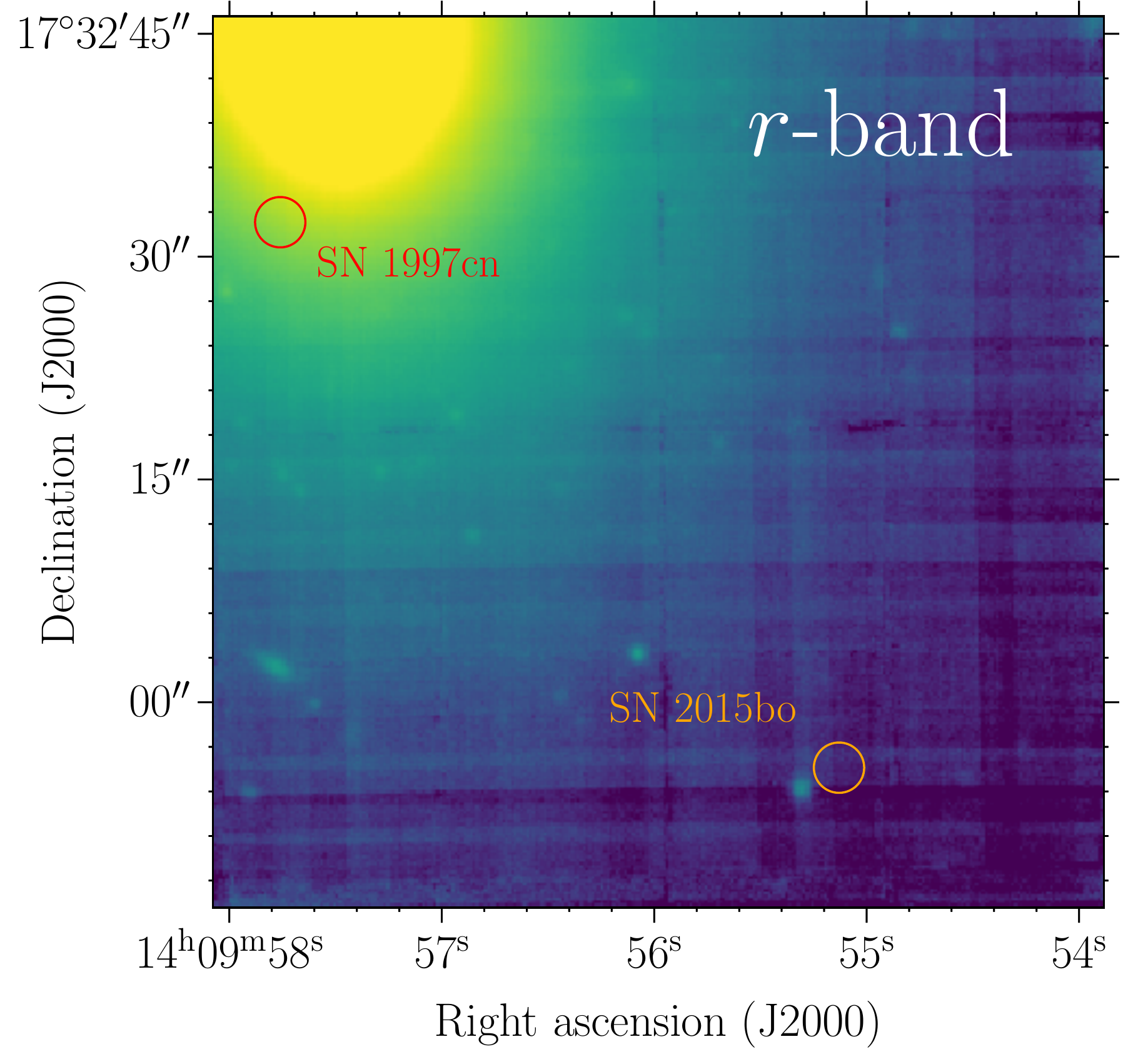}
\caption{The MUSE data-cube synthetic $r$-band image. The image was obtained by convoluting the spectra by a $r$ filter transmission. The orange and red circles represent a one square kpc aperture centered on the positions of \bo\ and \cn, respectively. \label{fig:muse}}
\end{figure}

Figure \ref{fig:muse} presents a synthetic $r$-band image extracted from MUSE observations of the host galaxy. \bo\ is distant from the center of the host galaxy, NGC 5490. This suggests that host-galaxy extinction is minimal for \bo. The object adjacent to \bo\ is an unrelated background galaxy.


\section{Photometry} \label{sec:phot}
\subsection{Light Curves} \label{sec:lc}
\begin{figure*}
\plotone{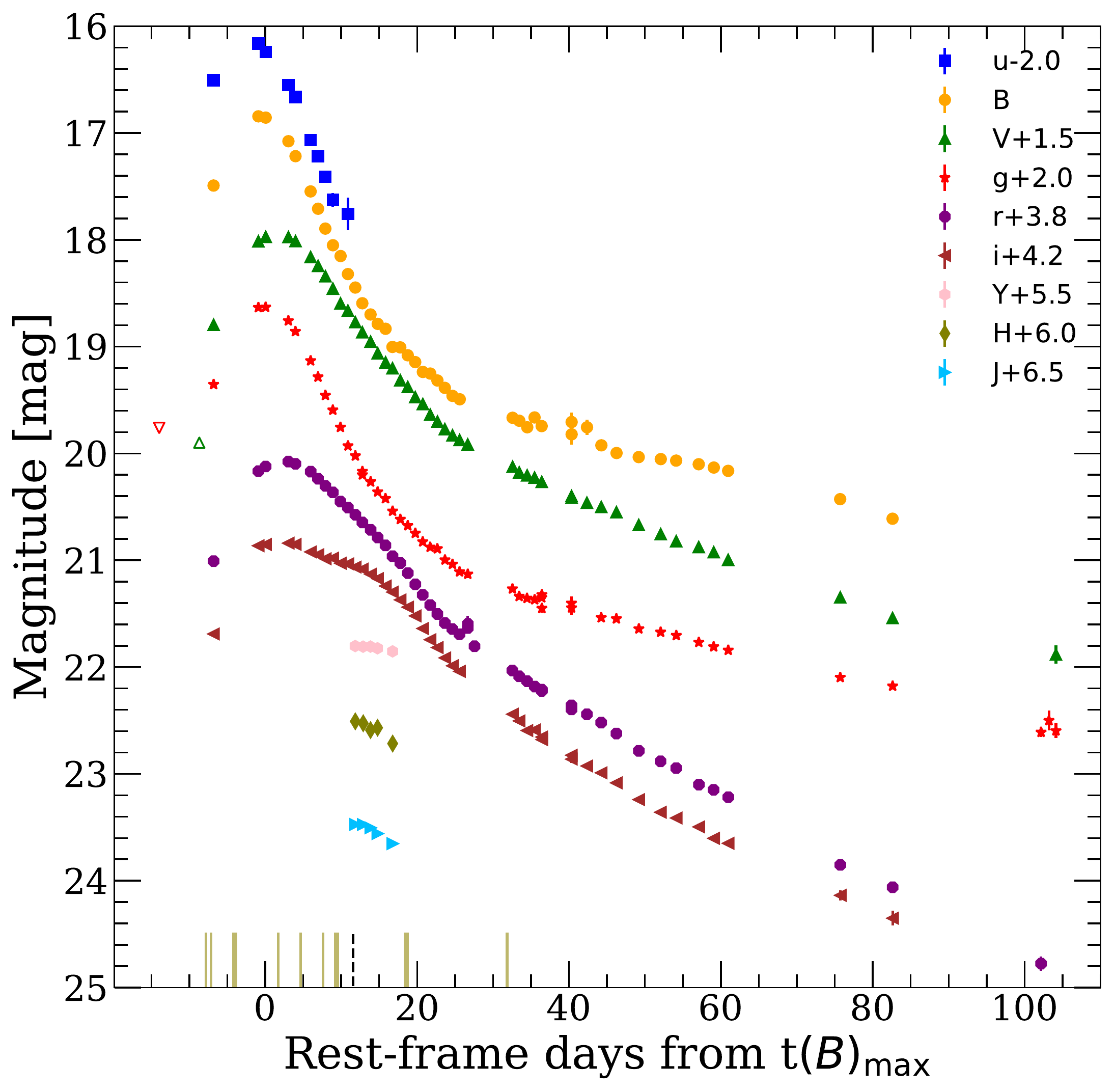}
\caption{Rest-frame optical and NIR photometry of SN~2015bo. The $V$-band discovery point is given as a green upward open triangle. The last-non detection from ASAS-SN in the $g$-band is provided as a red downward open triangle. Solid, dark khaki lines on the bottom horizontal axis represent epochs with optical spectra, and the dashed black line represents the epoch of our NIR spectrum. \label{fig:lc}}
\end{figure*}

The multi-band $uBVgriJHY$ light curves are presented in Figure \ref{fig:lc}. The filters are organized top-down from blue wavelengths to red wavelengths. Data from five optical filters ($BVgri$) extend to $\sim$+80~d after the time of $B$-band maximum (\timeB). $BVg$ rise on time scales longer than $\sim$10~d, but the true time of explosion is uncertain as the last non-detections are not constraining. 

To derive important photometric quantities such as \timeB, maximum m$_B$, and \DmB, we utilize the SuperNovae in object-oriented Python (SNooPy; \citealt{Burns11,Burns14}) template light-curve fitting software package\footnote{https://csp.obs.carnegiescience.edu/data/snpy}. The $B$ band was interpolated using a Gaussian process technique which underwent 150 Monte Carlo iterations to determine the uncertainties. The 1-$\sigma$ deviation from the Monte Carlo iterations was used as the uncertainty for the derived values. We used the 1991bg-like SN spectral energy distribution (SED) template within SNooPy\footnote{Taken from https://c3.lbl.gov/nugent/nugent$\_$templates.html.} to perform K-corrections. This correction template is used throughout this work for fitting fast decliners with \snpy. Light curve parameters of \timeB$=2457078.3 \pm 0.1$~d and maximum m$_B=16.83\pm 0.01$~mag were computed. \DmB\ and \sBV\ were calculated using the same direct Gaussian process interpolation technique. To calculate \sBV, only data with both $B$- and $V$-band taken on the same night were used. The values of \DmB\ and \sBV\ are $1.91 \pm 0.01$~mag and $0.48 \pm 0.01$, respectively. \nothing{Derived maximum apparent magnitudes in each CSP filter and their corresponding times relative to \timeB\ are presented in Table \ref{tab:max-app}. The $u$-band peaks $\sim$~2~days prior to \timeB, whereas the $V$-, $g$-, $r$-, and $i$- bands all peak after \timeB.}
\nothing{Throughout this work we use the parameters derived from the Gaussian process interpolation unless otherwise stated.}

\nothing{To assess if \bo\ is a `normal' 1991bg-like SNe, we fit the multi-band light curves with the EBV2 model in \snpy. Interestingly, the fits determine $E(B-V)_{host}=0.21 \pm 0.01$~(stat)~$\pm 0.06$~(sys). However, no narrow \NaI\ lines were seen in the spectra. An upper limit on the on \NaI\ pEW was calculated using the method described in \citet{Ashall21}. Using the method of \citet{Poznanski12} between \NaI\ D psuedo-equivalent width and $E(B-V)$ produced an upper limit of $E(B-V)_{host}=0.02 \pm 0.02$~mag, where the uncertainty is determined via the method of \citet{Phillips13}. Additionally, \bo\ is far away from its elliptical host galaxy, reducing the chance of significant host-galaxy extinction. We conclude that the host-galaxy extinction must be minimal, and that the SNooPY fits are inaccurate in this area of the parameter space. Therefore, throughout this work we opt to not correct for host-galaxy extinction.} 

\begin{figure}
\plotone{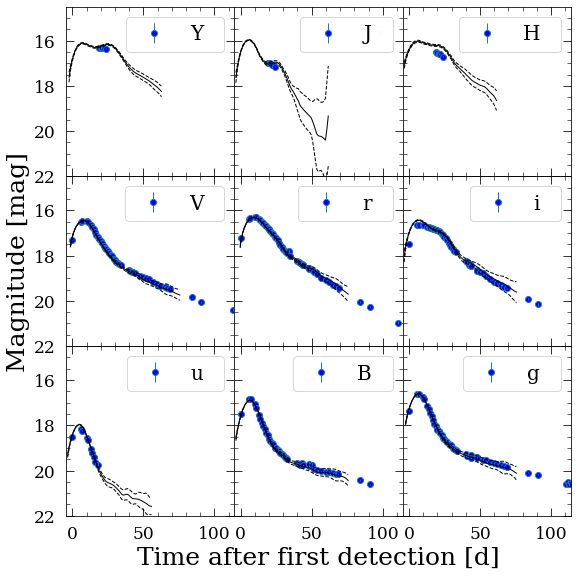}
\caption{SNooPy fits described in §\ref{sec:lc}. \nothing{These fits are produced with the EBV2 model, and the prior that the host-galaxy extinction is 0~mag.} The vertical axis is plotted in magnitudes, and the horizontal axis is in days after the first data point. \label{fig:snpy}}
\end{figure}

To determine the distance modulus to \bo, we use SNooPy EBV2 model \nothing{with the prior that the host-galaxy extinction, $E(B-V)_{host}~=~0$~mag}.  A distance modulus of $\mu$~=~34.33~$\pm$~0.01~(stat)~$\pm$~0.11~(sys)~mag was derived. Further discussion about distance calculations and a more accurate derivation of the distance modulus and its systematic uncertainty can be found in §\ref{sec:dist}. The \snpy\ fits (see Figure \ref{fig:snpy}) show that SN~2015bo's photometry is well matched by \bg\ light-curve templates.

\begin{deluxetable}{ccccc}
\tablenum{2}
\tablecaption{\nothing{Table of maximum light apparent magnitudes.} }\label{tab:max-app}
\tablewidth{0pt}
\tablehead{
\colhead{Filter} & \colhead{Magnitude} & \colhead{$\sigma$Magnitude} & \colhead{t$_{max}$ - \timeB} & \colhead{$\sigma$t$_{max}$}\\  \colhead{CSP-II} & \colhead{[mag]} & \colhead{[mag]} & \colhead{[days]} & \colhead{[days]} }

\startdata
$u$ & 18.15 & 0.01 & -1.9 & 0.3\\
$B$ & 16.44 & 0.01 &  0.0 & 0.1\\
$V$ & 16.84 & 0.01 &  2.0 & 0.1\\
$g$ & 16.62 & 0.01 &  0.6 & 0.1\\
$r$ & 16.27 & 0.01 &  2.7 & 0.1\\
$i$ & 16.62 & 0.01 &  2.1 & 0.1\\
$H$\tablenotemark{a} & 16.51 & 0.04 & 12.9 & 1.3 \\
\enddata
\tablenotemark{a}{The $H$ band was the only NIR band where \snpy\ was able to derive a maximum value. This is likely due to the lack of data in the NIR regime.}
\end{deluxetable}

\begin{figure*}
\plotone{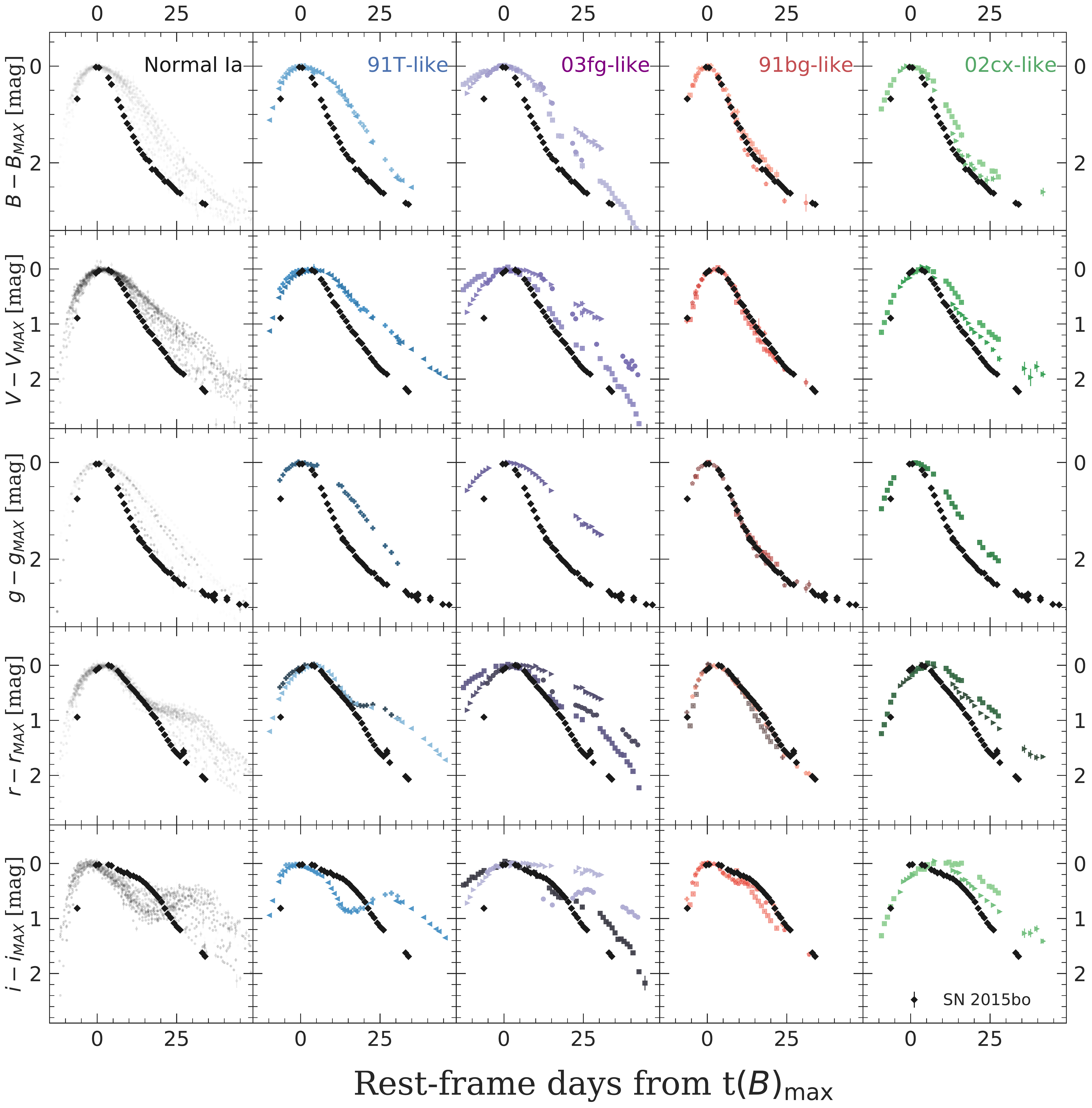}
\caption{The $B$-band (first row), $V$-band (second row), $g$-band (third row), $r$-band (fourth row), and $i$-band (fifth row) light curves compared to a selection of SNe Ia sub-types: normal (first column/grey), 91T-like (second column/blue), 03fg-like (third column/purple), 91bg-like (fourth column/orange), and 02cx-like (fifth column/green). \bo\ (black points) deviates significantly from all other types of \Ia, but matches quite well with other \bg. Comparison SNe are from \citet{Ashall20}. \label{fig:lc_comp}}
\end{figure*}

Figure \ref{fig:lc_comp} compares the light curves of SN~2015bo in $BVgri$ with a selection of sub-types of  \Ia\ from CSP-II. \bo's light curve resembles \bg, and it evolves more quickly than the other types of \Ia\ in every band \nothing{except the $i$-band}. The lack of a distinctive knee-type feature in the $r$-band when compared to normal and 1991T-like SNe highlights this difference. The \nothing{broader} $i$-band can be used to confirm if a SN is a sub-luminous SN. \bo, like many other sub-luminous SNe, lacks a secondary $i$-band maximum, whereas  normal and 1991T-like SNe have a distinct secondary maximum. In normal \Ia, the secondary $i$-band maximum comes from recombination of iron group elements \citep{Hoeflich02,Kasen06,Jack15}, however, in \bg, \citet{Kasen06} and \citet{Blondin15} propose the recombination either does not occur or occurs earlier which causes the maxima to merge and appear as one maximum in the light curve. This is also why the $i$-band peak occurs after $B$-band maximum. The importance of the second $i$-band maximum for using \Ia\ as distance indicators is discussed in §\ref{sec:dist}. The $i$-band for \bo\ is also slightly broader than the \bg\ comparison sample. Interestingly, \bo\ shows some similarities with 2002cx-like SNe in the $Bgr$ bands, but \bo\ declines significantly faster than 2002cx-like SNe in the $Vi$ bands. Although  2003fg-like SNe show a somewhat wider variety in their light curve morphologies, it is still evident that \bo\  declines more quickly than all 2003fg-like SNe in all plotted filters, and for most 2003fg-like SNe this difference in decline rate is significant. 

\begin{figure}
\plotone{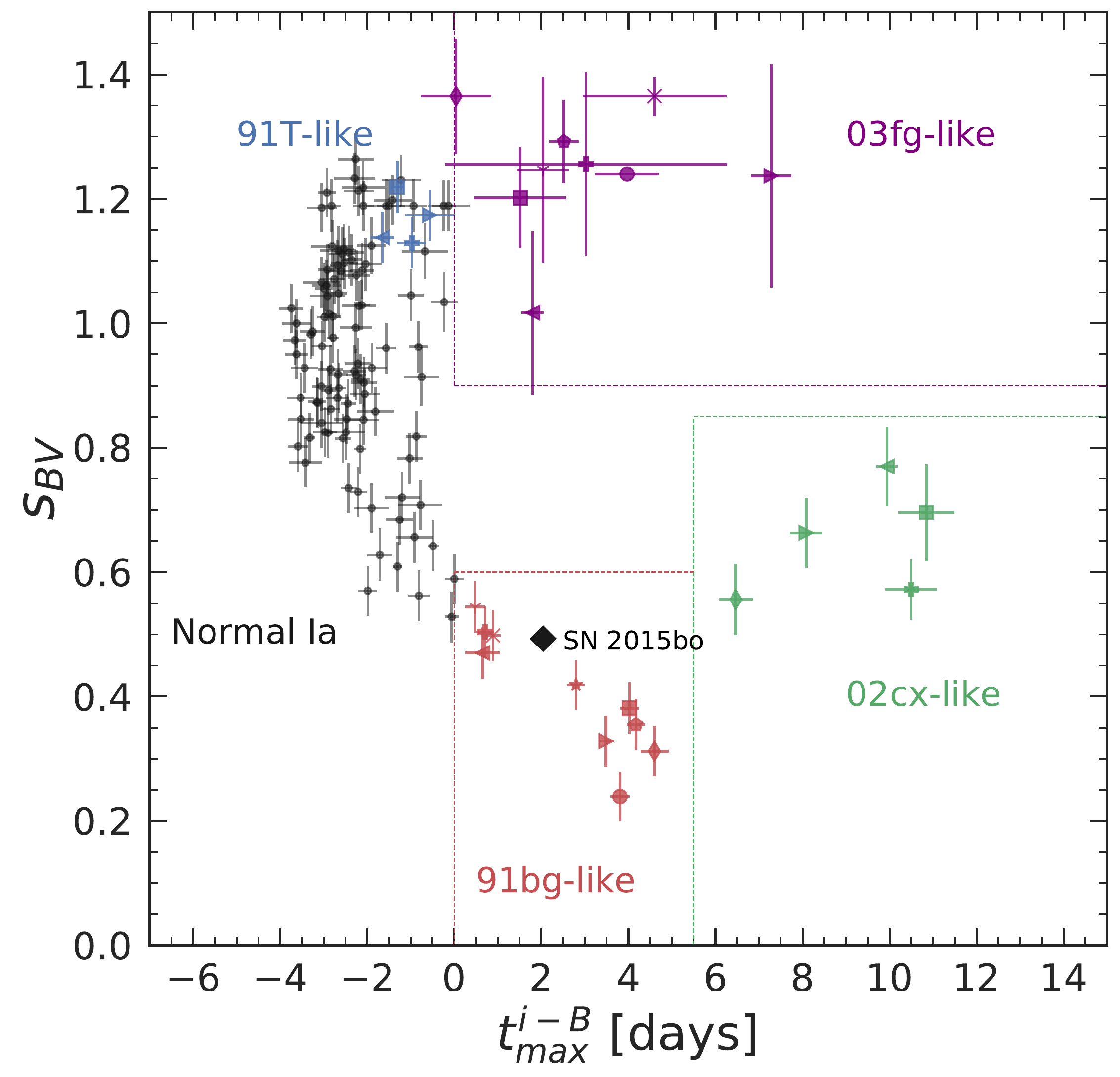}
\caption{The color stretch parameter s$_{BV}$ as a function of time difference between $i$- and $B$-band maxima, $t^{i-B}_{max}$. Different sub-types of \Ia\ occupy different areas of the plot, as shown in \citet{Ashall20}. \label{fig:sbv_vs_tiB}}
\end{figure}

\citet{Ashall20} showed SNe~Ia can be photometrically sub-typed by just their \sBV\ value and their time of primary  $i$-band maximum relative to $B$-band maximum (\iBmax). Figure \ref{fig:sbv_vs_tiB} presents this relationship with \bo\ overlaid. \bo\ is located in the \bg\ area of the parameter space where there are fewer SNe. This suggests the classification that \bo\ is a 1991bg-like SN. 

\subsection{Color Curves} \label{sec:cc}
\begin{figure}
\plotone{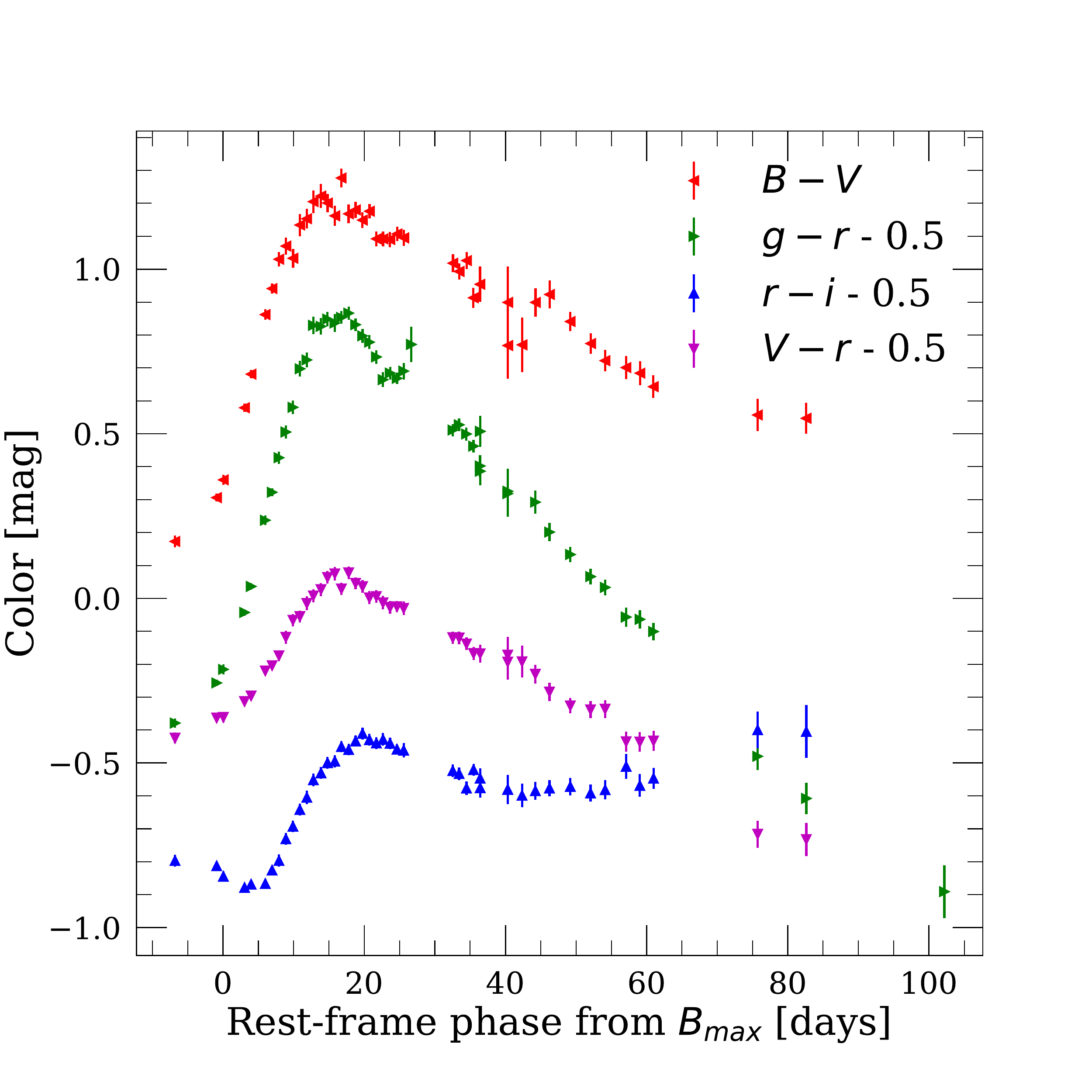}
\caption{Color versus rest-frame days from $B_{max}$ for \bo. From top to bottom, plotted colors are $B-V$ (red), $g-r$ (green), $V-r$ (magenta), and $r-i$ (blue). Color curves have been corrected for Milky Way, but not host-galaxy, extinction. \label{fig:cc} }
\end{figure}

In Figure \ref{fig:cc}, the color curves of \bo\ are presented. Throughout this work, we use K-corrected intrinsic colors. Color curves are obtained by subtracting photometric points observed on the same night. No interpolations between data points were performed.  At early times, the color is a diagnostic of the temperature of the photosphere, so Figure \ref{fig:cc} provides an approximate measurement of the temperature over time. Both the $B-V$ and the $g-r$ color curves are similar; they exhibit sharp increases from bluer to redder from $\sim$0~d to $\sim$+15~d and then turn to become bluer after $\sim$+15~d, however, $g-r$ declines more quickly than $B-V$ over this time frame. After $\sim$+55~d, the photosphere is well within the ejecta and the color no longer measures the temperature, but rather differences in ionization state from a mixture of absorption and emission features. At later epochs, $B-V$ flattens off to eventually reach a value of $\sim$0.5~mag at $\sim$+80~d, whereas $g-r$ continues to decline as far out as $\sim$+100~d to reach a value of $\sim-$0.4~mag. The $V-r$ color curve also peaks near $/sim$+15~d, but it does not rise as steeply as $B-V$ or $g-r$. When it declines, $V-r$ is broader than $g-r$ and is approximately the same slope as $B-V$. The $r-i$ color curve displays a slightly later red peak at around +20~d rather than around +15~d, and it also has a noticeably less steep increase. It also flattens off over time rather than decline. The flattening starts at $\sim$+30~d, which is earlier than when the $B-V$ color curve may start to flatten in the last two $B-V$ data points. These color curve shapes are abnormal; they may provide a potential explanation to why the  SNooPy \nothing{EBV2 model} wants to have a larger extinction value than can be derived from the  \nothing{\NaI\ D equivalent width}.

\begin{figure*}
\plotone{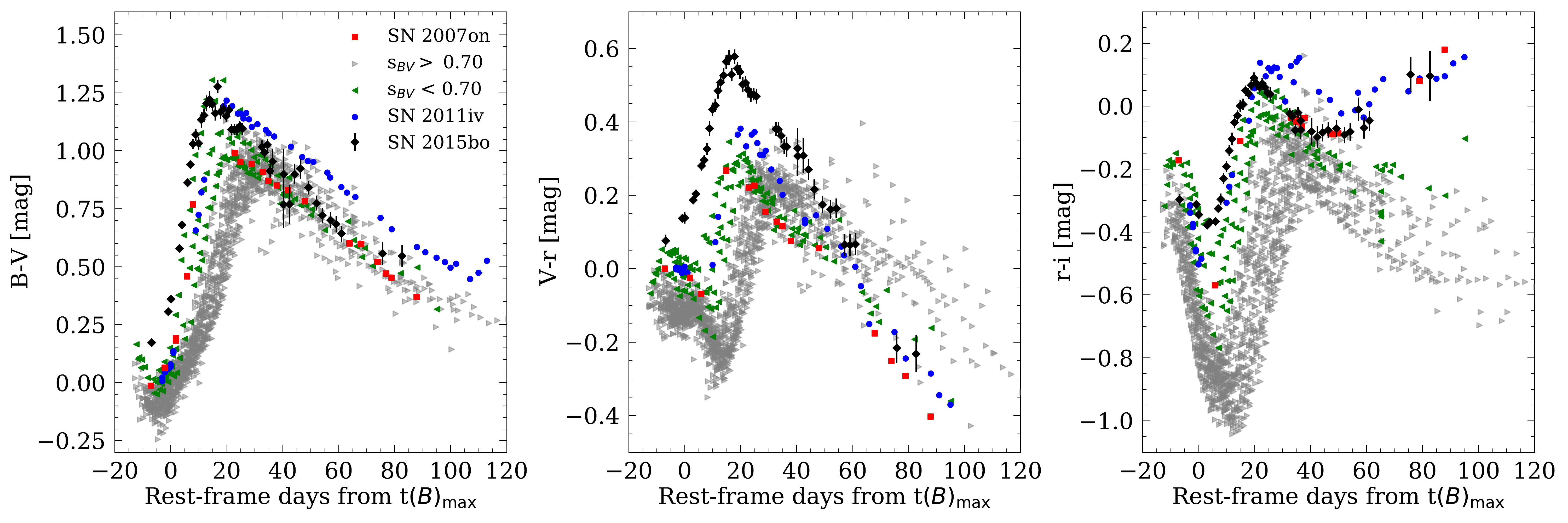}
\caption{Comparison of intrinsic color curves for \bo\ and other SNe. Plotted colors from left to right are $B-V$, $V-r$, and $r-i$. \bo\ is plotted in black diamonds with two similar \Ia, SN~2007on and SN~2011iv, plotted in blue circles and red squares, respectively. Green, left-facing triangles are other fast decliners; the condition \sBV $< 0.70$ effectively segregates between normal and fast-declining \Ia. Grey, right-facing triangles are all other types of \Ia.  All comparison \Ia\ data points have been corrected for both Milky Way and host-galaxy extinction, whereas \bo\ is only corrected for galactic extinction since there is most likely an insignificant amount of host-galaxy extinction. \label{fig:cc-comp}}
\end{figure*}

Figure \ref{fig:cc-comp} compares the intrinsic color curves of \bo\ to a variety of other \Ia. The comparison \Ia\ are corrected for both their respective galactic and host-galaxy extinctions. \bo\ is plotted in black diamonds, with two transitional \Ia, SN~2007on and SN~2011iv, plotted in blue circles and red squares, respectively. These two transitional objects are plotted individually because they also exhibit the same peculiar blue rise in the $r-i$ color curve that \bo\ has after $\sim$+40~d (see below for details). SN~2007on and SN~2011iv are interesting objects in their own right; they both exploded in NGC 1404, and they are both transitional \Ia. Detailed discussions of SN~2007on and SN~2011iv can be found in \citet{Gall18} and \citet{Ashall18}.

In $B-V$, \bo\ starts bluer than the comparison sample and becomes more red that the comparison sample over time after $\sim$+30~d. Its Lira tail (the region in the $B-V$ color curve after $\sim$+20~d for which the majority of SNe have similar slopes; \citealt{Phillips99}) is the steepest of our sample. Comparing SN~2007on and SN~2011iv, \citet{Gall18} found SN~2011iv was bluer earlier on, yet it became redder than SN~2007on at $\sim$+30~days. \citet{Gall18} interpreted this as an effect from the progenitor white dwarfs having differing central densities at the time of explosion. Therefore, the \Nifs\ distributions would be different for SN~2007on and SN~2011iv. At $B$-band maximum and at the reddest color inflection point, \bo\ has a similar $B-V$ color to SN~2007on; however, the slope of the decline in the Lira tail is steeper in \bo\ than both SN~2007on and SN~2011iv. This could be an indication that the central density of \bo\ is even lower than SN~2007on. 

The $V-r$ color curve for \bo\ is unique. It reaches a red peak higher than any other observed SN Ia in our comparison sample. This is peculiar, especially considering this same behavior does not occur in either $r-i$ or $B-V$ colors, or with other transitional or \bg. Additionally, it seems that the triumvirate of SN~2007on, SN~2011iv, and \bo\ decline in $V-r$ more rapidly than normal \Ia\ after $\sim$+50~d past \timeB. 

Another intriguing property of SN~2007on, SN~2011iv, and \bo\ is the redward rise of the  $r-i$ color curve at $\sim$+40~d after $B$-band maximum. This red color shift could be due to the fact that fast decliners are [\CaII]  strong. Therefore the [\CaII] $\lambda \lambda$7291.5, 7323.9 emission, which is usually seen beginning at $\sim$+40~d \citep{Taubenberger17}, will affect the $r-i$ color curve. This difference in the $r-i$ color curve after $\sim$+40~d could provide a robust way to photometrically distinguish a subset of transitional and/or  \bg\ that may share a progenitor scenario or explosion mechanism or to distinguish fast-declining SNe from normal \Ia. 

The $r-i$ color minimum that occurs just after \timeB\ for SN~2007on, SN~2011iv, and \bo\ are also significantly earlier and bluer than normal \Ia. This earlier first turning point could provide a good way to identify sub-luminous \Ia\ in future studies. We see no evidence for an inflection point at $\sim$0-10~d in the $V-r$ color despite having data during the epochs where all other \Ia\ in the sample turn.


\subsection{Luminosity-Width Relation}\label{sec:LWR}

The peak absolute magnitude of \bo\ is calculated from the light curves to be $M_B$~=~\nothing{$-17.67$}~$\pm$~\nothing{0.15}~mag. This value has been corrected for only foreground extinction \nothing{and not host-galaxy extinction.} This absolute magnitude is derived using the Cosmic Microwave Background (CMB) corrected distance modulus, which is $34.21\pm0.15$~mag.

\begin{figure*}
\plotone{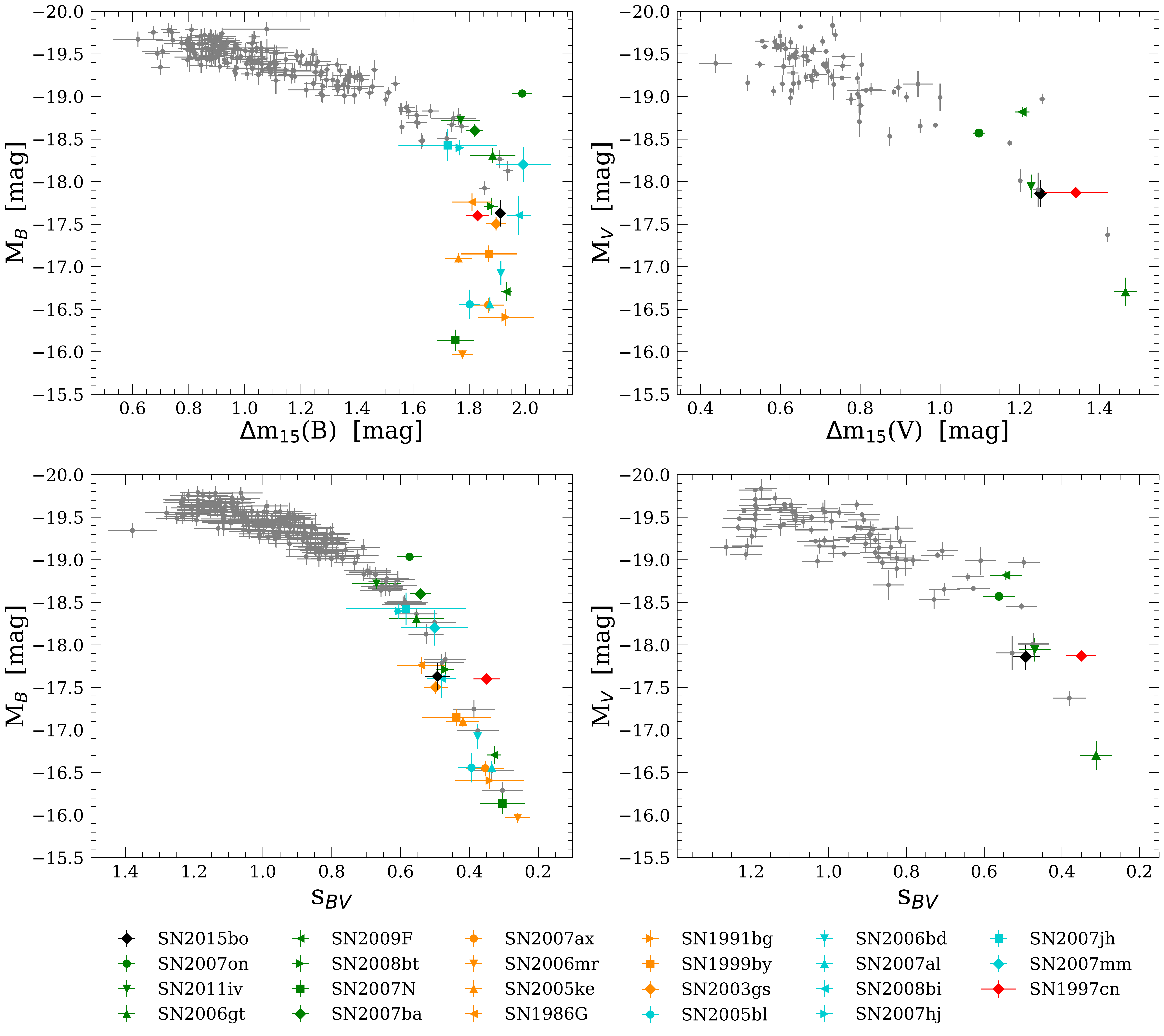}
\caption{Luminosity-width relation using \sBV, \DmB, and \DmV. For all panels, the \nothing{black diamond} is \bo. \nothing{The absolute magnitude for \bo\ was computed without host-galaxy extinction (i.e., $E(B-V)_{host} = 0.00$). The \sBV\ and peak magnitude for \bo\ were both calculated with a direct Gaussian interpolation}. Other fast decliners are plotted as green, orange, or cyan diamonds, triangles, squares. \cn\ is plotted as the \nothing{red} diamond. Grey points are a sample of other sub-types and normal \Ia. \emph{Top-Left}: Absolute $B$-band magnitude plotted against \DmB. \emph{Bottom-Left}: Absolute $B$-band magnitude plotted against \sBV. \emph{Top-Right}: Absolute $V$-band magnitude plotted against \DmV. \emph{Bottom-Right}: Absolute $V$-band magnitude plotted against s$_{BV}$. 
For the background \Ia\ on the left-hand side, \sBV\ and \DmB\ values come from \nothing{the EBV2 model within \snpy}, whereas on the right hand side \sBV\ and \DmV\ values for the background \Ia\ are directly computed by \snpy\ \nothing{using a direct Gaussian interpolation} on the specific band. 
All background SNe satisfy the following: $z > 0.01$, $A_V < 0.75$~mag, $\sigma M_V < 0.3$~mag, $\sigma {\Delta \mathrm{m}_{15}\mathrm{(V)}} < 0.05$~mag, and $\sigma {s_{BV}} < 0.05$. These choices were made to eliminate poorly measured background SNe or objects that suffered from significant extinction and/or uncertain distances due to peculiar velocities. \label{fig:phil-combo}}
\end{figure*}

The luminosity-width relation is plotted in Figure \ref{fig:phil-combo} using comparison SNe from the CSP-I and CSP-II. \bo\ is plotted without corrections for host-galaxy extinction, and the rest of the sample is corrected for host galaxy-extinction. The left side of the figure is for the $B$-band, whereas the right side of the figure is for the $V$-band. \bo\ is located with the transitional and \bg\ and is within the area of parameter space of \sBV$\approx$0.5, where there is a lack of objects \citep{Ashall16b}. \nothing{The background comparison SNe in} the left panels of the figure use absolute magnitude values from SNooPy template fits \nothing{using the EBV2 model}; \nothing{the comparison SNe in } the background comparison SNe in the right panels use $M_V$ and \DmV\ from direct Gaussian process fits with the same method as in §\ref{sec:lc}. This explains the difference in scatter between the luminosity-width relation in the $B$-band compared to the $V$-band. 

In the top left panel, \bo\ is located in the degenerate region of the luminosity-width relation parameter space. This degeneracy arises from \DmB\ obtaining the same value for different light curve shapes \citep{Burns14}. However, when the correlations are plotted as a function of \sBV, \bo\ occupies the bright end of the \bg\ parameter space in the luminosity width relation. In the $B$ band, $M_B$ is among the parameter space occupied by other fast decliners. In the $V$ band, \bo\ is in the fast-declining, under-luminous parameter space as well. 

SN~1997cn is also plotted in Figure \ref{fig:phil-combo} using the photometry from \citet{jha06}. Due to the photometry starting well after maximum in all bands, we were unable to directly calculate the time of $B$- or $V$-band maxima, maximum magnitudes in $B$ and $V$ bands, \DmB, \DmV, or \sBV . Because of this, values from the best-fit templates derived from the SNooPy EBV2 model were used for \cn. 
In all panels, \cn\ occupies the same parameter space as \bo. 


\subsection{Bolometric Light Curves}\label{sec:bolo}
Pseudo-bolometric light curves were constructed using the
direct spectral energy distribution method within SNooPy. The wavelength range was from 3,000~\AA\ to 19,370~\AA. In short, the method converts the magnitudes into fluxes to produce a spectral energy distribution (SED), the 1991bg-like spectral templates are used for K- and S-corrections, and the SED is integrated and scaled with respect to the distance modulus to produce a bolometric flux at each epoch. Due to the lack of temporal coverage in the infrared data, we use a Rayleigh-Jeans model to extrapolate our SEDs when data was not available. The SEDs were also corrected for Milky Way extinction. The error on the bolometric light curve was determined via varying the distance modulus through 100 Monte Carlo iterations. To do this, the distance modulus was treated as a Gaussian distribution. A Gaussian process was used to determine the maximum for each iteration. The mean value of these maxima from the Monte Carlo iterations was taken as the final peak pseudo-bolometric luminosity and the mean of the standard deviations of the maxima was taken as the uncertainty.

\begin{figure}
\plotone{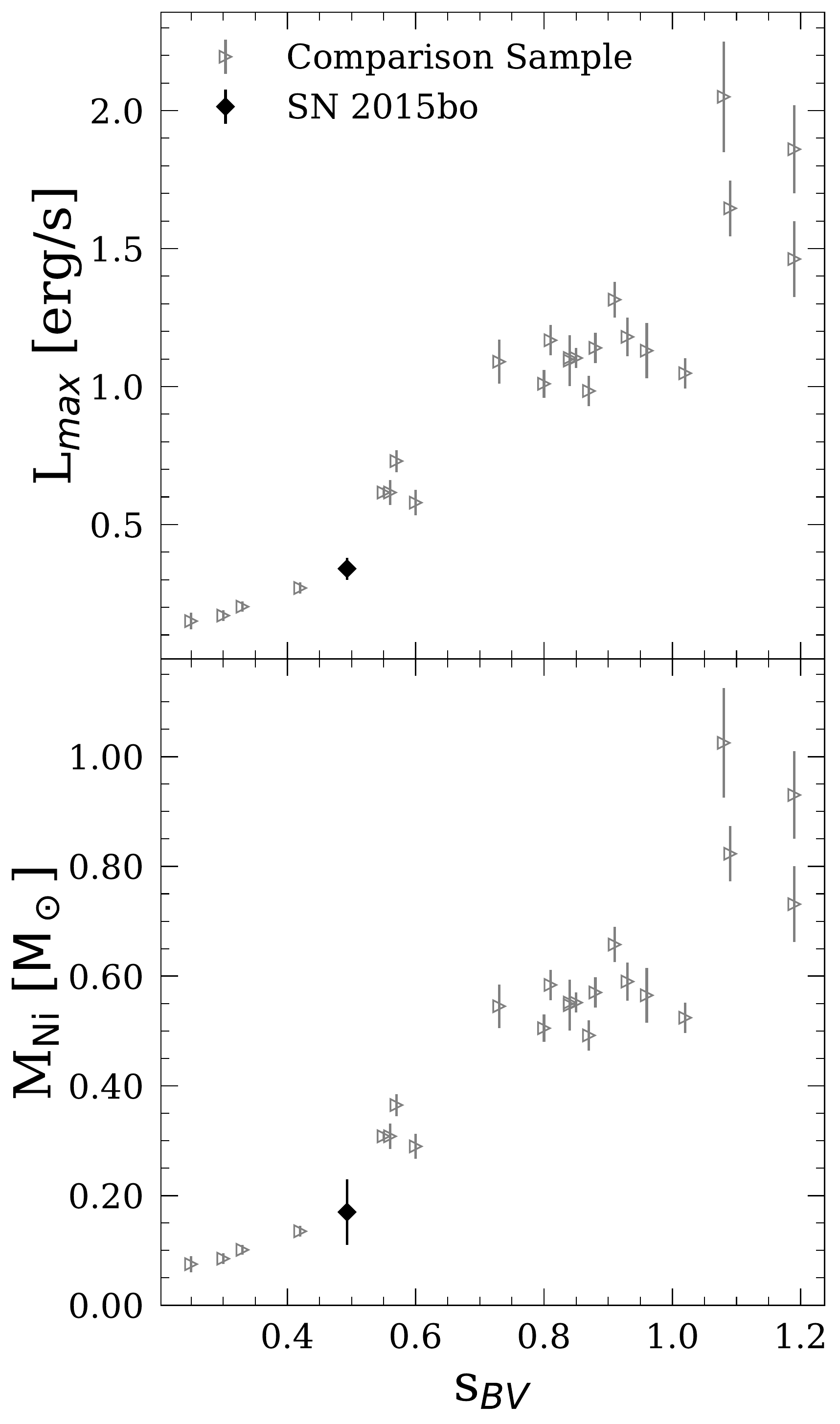}
\caption{Pseudo-bolometric luminosity (top) and \Nifs\ mass (bottom) as a function of \sBV. \bo\ is plotted as the \nothing{black diamond}, while other \Ia\ are \nothing{hollow, grey right-facing triangles}. \label{fig:bolo}}
\end{figure}

The peak pseudo-bolometric luminosity of \bo\ is $L^{peak}_{bolo}=0.34 \pm 0.04 \times 10^{43}$~\ergs. Using the relation between $L^{peak}_{bolo}$ and \Nifs\ mass presented in \citet{Stritzinger05} and \citet{Arnett82}, a \Nifs\ mass of $0.17 \pm 0.06 {M_\odot}$ is derived. Calculated values are compared to other \Ia\ presented in \citet{Ashall18} in Figure \ref{fig:bolo}. Both the bolometric luminosity and \Nifs\ mass of \bo\ are consistent with the general trend presented in \citet{Ashall19a}. \bo\ has a lower luminosity and lower \Nifs\ mass than the majority of the CSP comparison sample. This places it near other \bg. 

A similar pseudo-bolometric light curve and \Nifs\ mass were not computed for \cn\ because the data for \cn\ is insufficient to directly derive a \Nifs\ mass. However, \citet{Turatto98} found a \Nifs\ mass of $\sim$0.1\Msun for \cn\ using spectral models.


\section{Spectra} \label{sec:spec}
Photometry is a powerful tool for classifying and studying supernovae, but it does not reveal the full picture. Spectroscopic observations complement photometric ones and enable further understanding of the elemental composition of SNe. Due to the homologous \nothing{($r\propto vt$)} expansion of a SN ejecta $\sim$10~s after the explosion, obtaining a time series of spectra allows the reconstruction of the evolution of the ejecta over time. In addition to providing composition information, spectra also can be used to derive velocities of the ejecta and provide additional information on the physics of the explosion such as the temperature, the ionization state, and the kinetic energy.

\begin{deluxetable*}{ccccc}
\tablenum{3}
\tablecaption{Log of spectroscopic observations.}\label{tab:spec}
\tablewidth{0pt}
\tablehead{
\colhead{UT Data} & \colhead{MJD} & \colhead{Epoch}\tablenotemark{a} &  \colhead{Telescope} & \colhead{Spectrograph} \\ & \colhead{[days]} & \colhead{[days]} & & }

\startdata
2015-02-15 &    57068.35       & $-$7.8   &	NTT\tablenotemark{b} &	EFOSC2\\
2015-02-16 &	57069.00       & $-$7.2   &	LT\tablenotemark{c} & SPRAT\\
2015-02-19 &	57072.00       & $-$4.2   &	LT & SPRAT\\
2015-02-19 &	57072.34       & $-$3.9   &	NTT	& EFOSC2\\
2015-02-19 &	57072.36       & $-$3.9   &	NTT	& EFOSC2\\
2015-02-25 &	57078.00       & 1.7    &	DUP\tablenotemark{d}&WFCCD \\
2015-02-28 &	57081.00       & 4.7    &  LT &	SPRAT\\
2015-03-03 &	57084.00       & 7.6	 &  LT & 	SPRAT \\
2015-03-03 &	57085.64	   & 9.2	 &  NOT\tablenotemark{e} & ALFOSC \\
2015-03-05 &	57086.00       & 9.6	 &  LT & SPRAT \\
2015-03-07 &	57088.00       & 11.5	 &  Magellan &	FIRE \\
2015-03-10 &	57092.69	   & 16.2	 &  NOT	&   ALFOSC \\
2015-03-14 &	57095.00       & 18.4   &  LT &	SPRAT \\
2015-03-14 &	57095.30       & 18.7   &	NTT &	EFOSC2\\
2015-03-14 &	57095.32       & 18.8   &	NTT &	EFOSC2\\
2015-03-26 &	57108.63	   & 31.8	 &  NOT	&   ALFOSC
\enddata
\tablenotemark{a}{Epoch phase relative to rest frame $B$-band maximum.} \\
\tablenotemark{b}{New Technology Telescope.} \\
\tablenotemark{c}{Liverpool Telescope.} \\
\tablenotemark{d}{du Pont Telescope.} \\ 
\tablenotemark{e}{Nordic Optical Telescope.}
\end{deluxetable*}

\subsection{Optical Spectra}

\begin{figure*}
\plotone{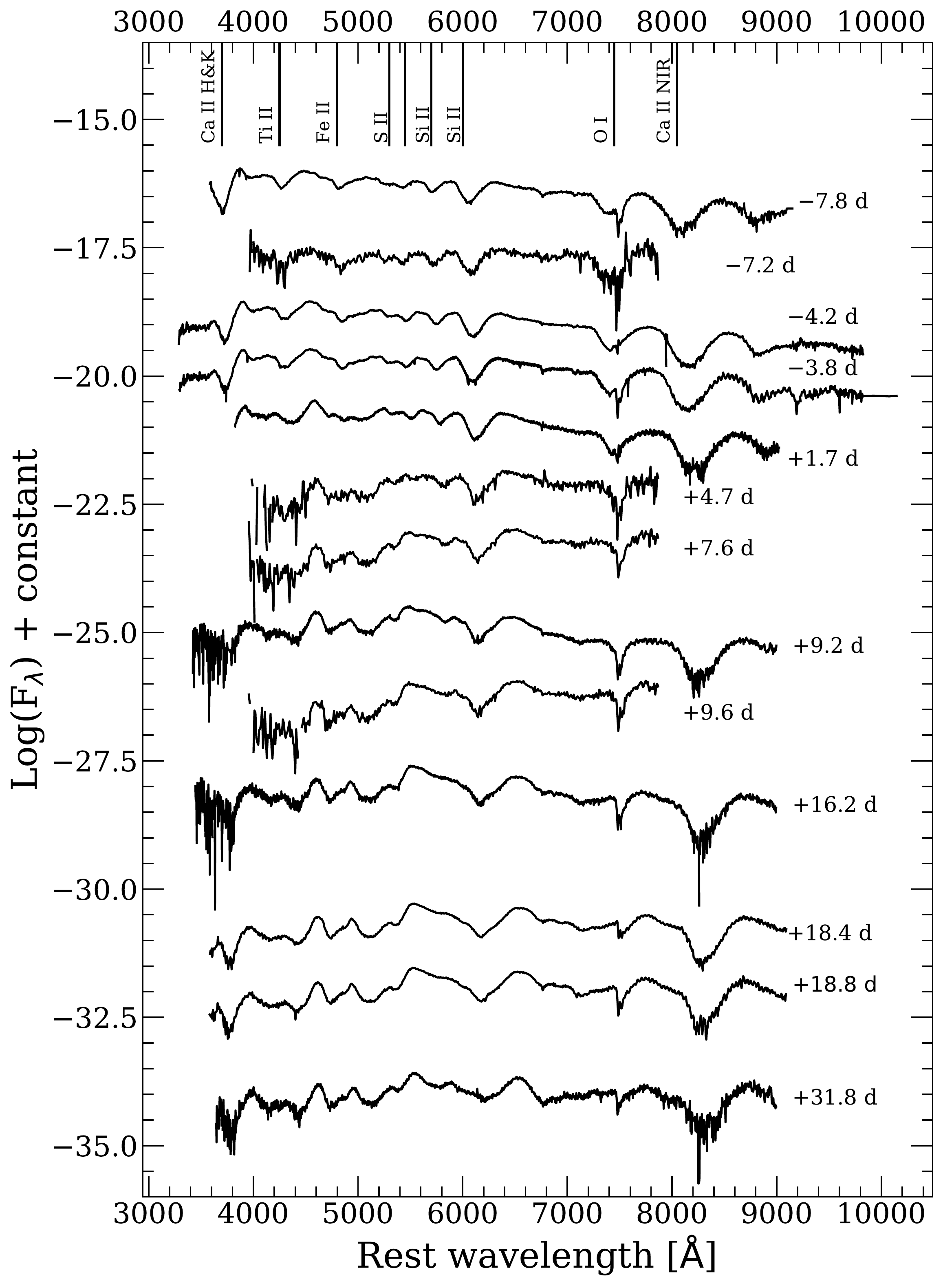}
\caption{Time series of optical spectra for SN~2015bo. Times relative to \timeB\ are provided next to each spectrum. \label{fig:spec}}
\end{figure*}

Figure \ref{fig:spec} presents a time series of the optical spectra compiled here for \bo. The spectra span from $-$7.8~d to +31.8~d relative to $B$-band maximum. The spectra are dominated by features of intermediate-mass and Fe-group elements as well as oxygen. The main spectral features we identify are: \CaII\ H\&K  $\lambda \lambda$3968, 3933, \TiII\ $\lambda$4395, \FeII\ $\lambda \lambda$4923, 5169, \FeIII\ $\lambda$5156, \SII\ $\lambda \lambda$5449, 5623, \SiII\ $\lambda \lambda$ 5972, 6355, \OI\ $\lambda$7774,  and \CaII\ $\lambda \lambda$8498, 8542, 8662 NIR lines. These lines are identified based on \citet{Mazzali97} and \citet{Ashall16b}.

\begin{figure*}
\plotone{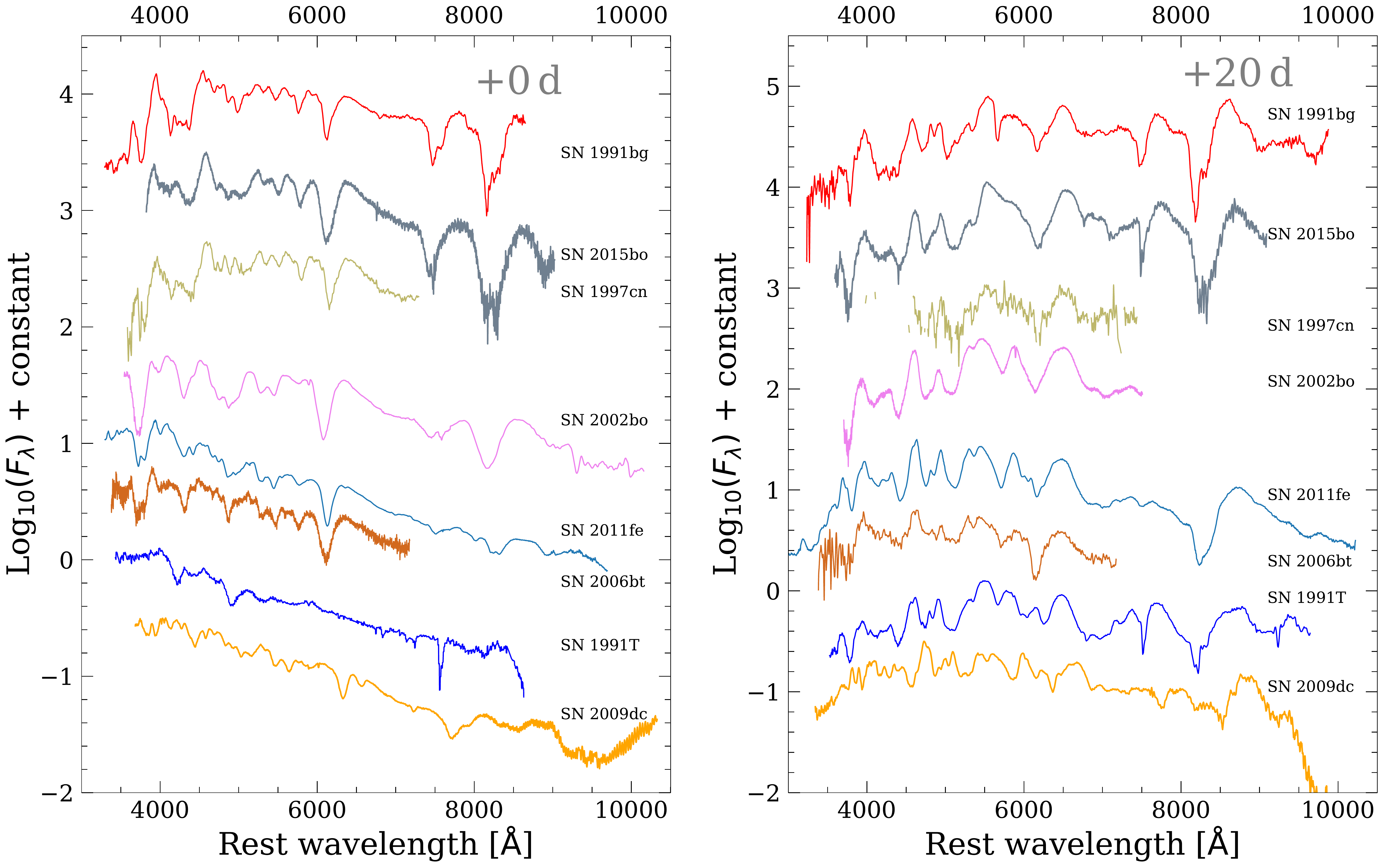}
\caption{Comparison of \bo\ (grey) to other \Ia : SN~1991bg \citep{fil92a}; normal \Ia\ SN~2002bo \citep{Benetti04,Szabo03} and  SN~2011fe \citep{Nugent11}; peculiar SN~2006bt \citep{Foley10}; the over-luminous SN~1991T \citep{phi92,fil92b}; and the 2003fg-like SN~2009dc \citep{Yamanaka09}. \emph{Left}: Comparison at \timeB. \emph{Right}: Comparison at +20~d from \timeB.  \label{fig:spec-comp-two}}
\end{figure*}

Spectra of \bo\ are compared to other sub-types of \Ia\ at maximum light and +20~d in Figure \ref{fig:spec-comp-two}. All objects in the sample are dominated by intermediate-mass elements and Fe-group elements. Most have a strong \SiII\ $\lambda$6355 feature. \bo\ is most similar to SN~1991bg, as demonstrated by the strong \SiII\ $\lambda$ 5972 feature. In other SNe, the \SiII\ $\lambda$5972 feature is weak or entirely absent (e.g. SN~1991T and SN~2011fe). While not as strong as SN~1991bg, \bo's \TiII\ $\lambda$4400 feature is still present. This feature (\TiII\ $\lambda$4400) is only observed in our sample in \bo, SN~1991bg and SN~2006bt. As the strength of the \SiII\ $\lambda$5972 and \TiII\ $\lambda$4400 is indicative of temperature, this suggests \bo\ has a photospheric temperature between those of SN~1991bg and a normal Ia such as SN~2011fe. By +20~d, all of the SNe in the sample look similar because the photosphere is well within the \Nifs\ region.

\begin{figure}
\plotone{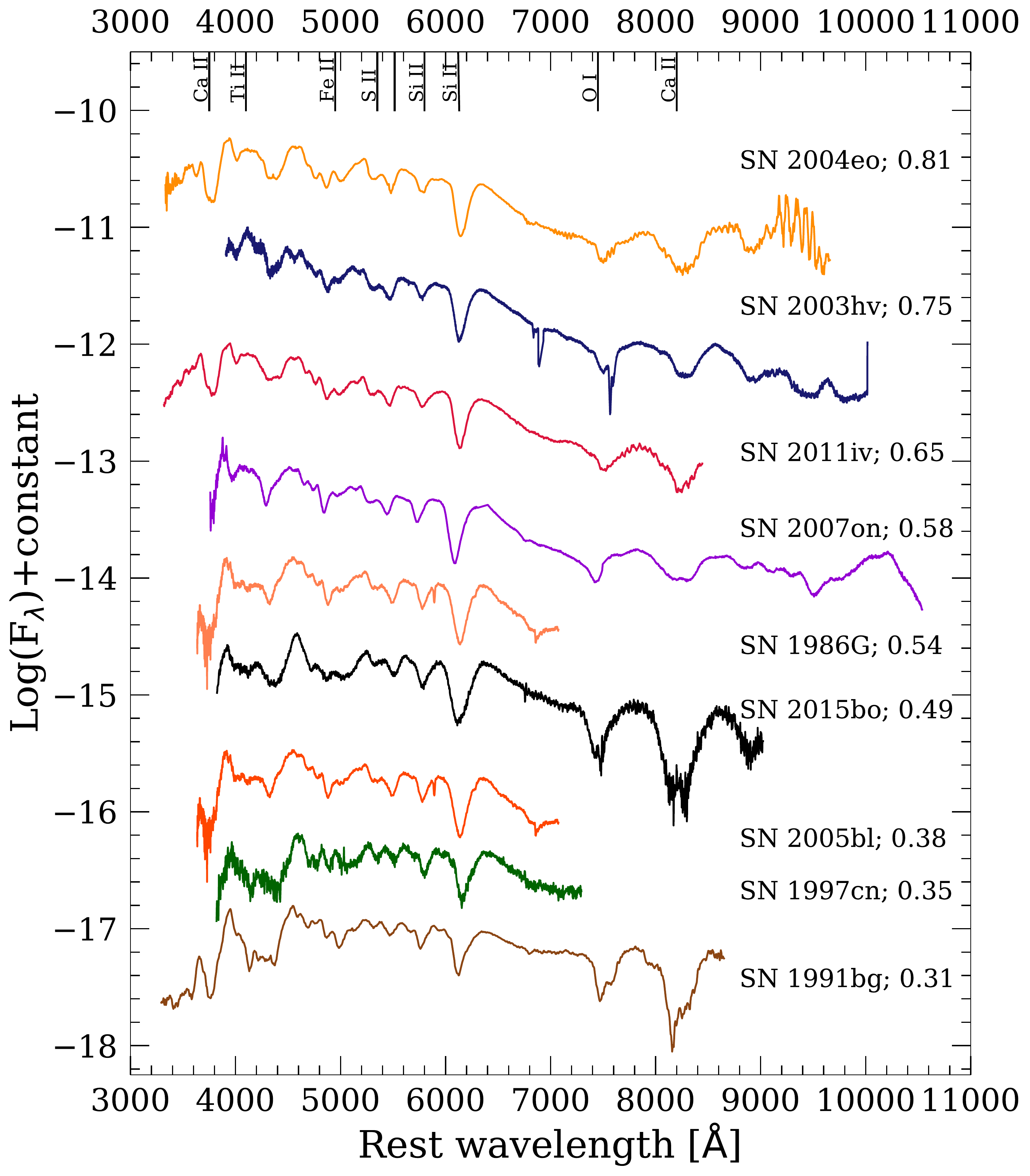}
\caption{Comparison of optical spectra at \timeB\ magnitude of 1991bg-like and transitional \Ia. From blue to red, the significant features are \CaII\ H\&K $\lambda \lambda$3968, 3933, \TiII\ $\lambda$4400, \FeII\ $\lambda \lambda$4923, 5169, \SII\ $\lambda \lambda$5449, 5623, \SiII\ $\lambda\lambda$5972, 635, \OI\ $\lambda$7774, and \CaII\ $\lambda \lambda$8498, 8542, 8662. \sBV\ values are taken from \citet{Ashall18} Table 5, except for SN~1991bg which is taken from \citet{Galbany19} Table 3.  \label{fig:spec-comp}}
\end{figure}

Figure \ref{fig:spec-comp} compares \bo\ to other SN~1991bg-like and transitional \Ia. The sample is arranged by \sBV\ value; there is a trend that SNe with smaller \sBV\ have stronger \SiII~$\lambda$5972 features. The \TiII~$\lambda$4400 feature also shows the same relationship between depth and \sBV. Within this trend, \bo\ is located between SN~2005bl and SN~1986G both in terms of \sBV\ and strength of the \TiII~$\lambda$4400 and \SiII~$\lambda$5972 features. Figures \ref{fig:phil-combo} and \ref{fig:bolo} confirm this; in both plots, \bo\ resides on the more luminous end of the fast decliners. These photometric results are consistent with spectral observations. \cn\ is also plotted to compare its spectra to the other SNe -- it falls between SN~2005bl and SN~1991bg based on \sBV\ value.

Discussing spectral differences in a qualitative manner is helpful to see general trends, however, as a scientific method it is imprecise. Therefore, velocities and pseudo-equivalent widths are derived to provide a quantitative picture of \bo\ and its spectral relationship to other \Ia. These measurements are presented in §\ref{sec:vel-and-pew}.

\subsection{Velocity and pEW Fitting Method}
We fit all optical spectra using the Measure Intricate Spectral Features In
Transient Spectra\footnote{https://github.com/sholmbo/misfits.} (mistfits; S. Holmbo et al., in prep.) fitting program. This program was used to calculate both the wavelengths of each spectral feature and the pseudo equivalent-widths (pEW). To fit the wavelengths, spectra were smoothed using misfits' low-pass fast Fourier transformed filter spectra following the method described in \citet{Marion09}. An uncertainty spectrum was computed from the difference between the observed and smoothed spectra. The absolute values of the residuals were smoothed using a Gaussian filter where the smoothed residuals contain 68\% of the absolute value of the residual level. This is used as the 1-$\sigma$ error spectrum. 

From there, a Gaussian function and a linear continuum were simultaneously fitted to each feature using the \emph{velocity.gaussians} function within misfits. The initial guess continuum boundary points were selected by hand, and the best fit was determined by $\chi ^2$ minimization. A final wavelength and uncertainty were respectively determined using the mean and standard deviation of 1000 Monte Carlo iterations. This uncertainty was added in quadrature with the uncertainty of the instrument resolution to arrive at the final wavelength uncertainty. The relativistic Doppler formula and rest wavelength of the feature were used to convert the wavelength of the absorption minimum into velocities.

Pseudo-Equivalent Width determination followed the procedure presented in \citet{Garavini07} using misfits' \emph{width.shallowpew} function. The mean and standard deviation of 1000 Monte Carlo iterations were used as the pEW value and $1$-$\sigma$ uncertainty. 

\subsection{Branch Diagram, Velocities, and pEW}\label{sec:vel-and-pew}
\subsubsection{Branch Diagram}\label{sec:branch}
\begin{figure}
\plotone{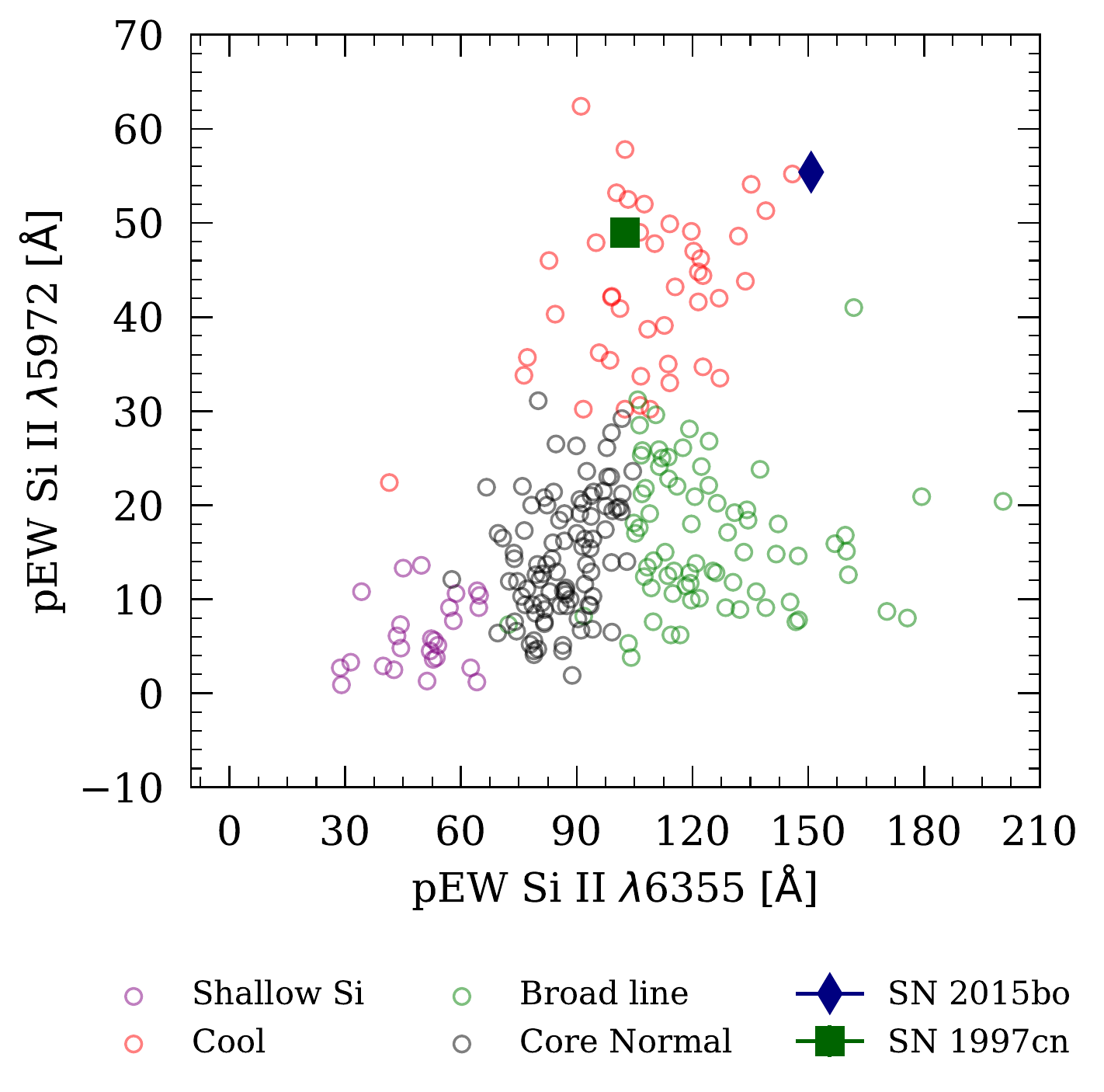}
\caption{Branch diagram \citep{Branch06} with SN~2015bo as the blue diamond and SN~1997cn as the dark green square. Data for other SNe Ia are from \citet{Blondin12} and \citet{Folatelli13}. \label{fig:branch}}
\end{figure}

An additional way to classify \Ia\ into sub-types is to plot the pEW of \SiII\ $\lambda$5972 against \SiII\ $\lambda$6355 at maximum \citep{Branch06}. Figure \ref{fig:branch} shows \bo\ along with other \Ia. There are four regions: Shallow Silicon, Core Normal, Cool, and Broad Line. Core Normal SNe are a highly homogeneous group with similar spectral features. Broad Line SNe are similar to Core Normal, but have broader features. Likewise, Shallow Silicon are also similar to Core Normal but tend to have equally weak \SiII\ $\lambda$5972 and \SiII\ $\lambda$6355 features. Shallow Silicon SNe tend to be over-luminous. Finally, there are the Cool SNe. These objects are sub-luminous -- strong \SiII\ $\lambda$5972 values are predominantly seen in sub-luminous SNe. \bo\ occupies the upper right area of the Cool region among other fast-declining, sub-luminous \Ia, whereas \cn\ falls in the upper central area of the Cool region. \nothing{While \bo\ belongs to the Cool SNe, it shows the bluest intrinsic $V-r$ color of our comparison sample of \Ia. This contradiction could imply that the luminosity does not come from the line forming region.}

\begin{figure*}
\plotone{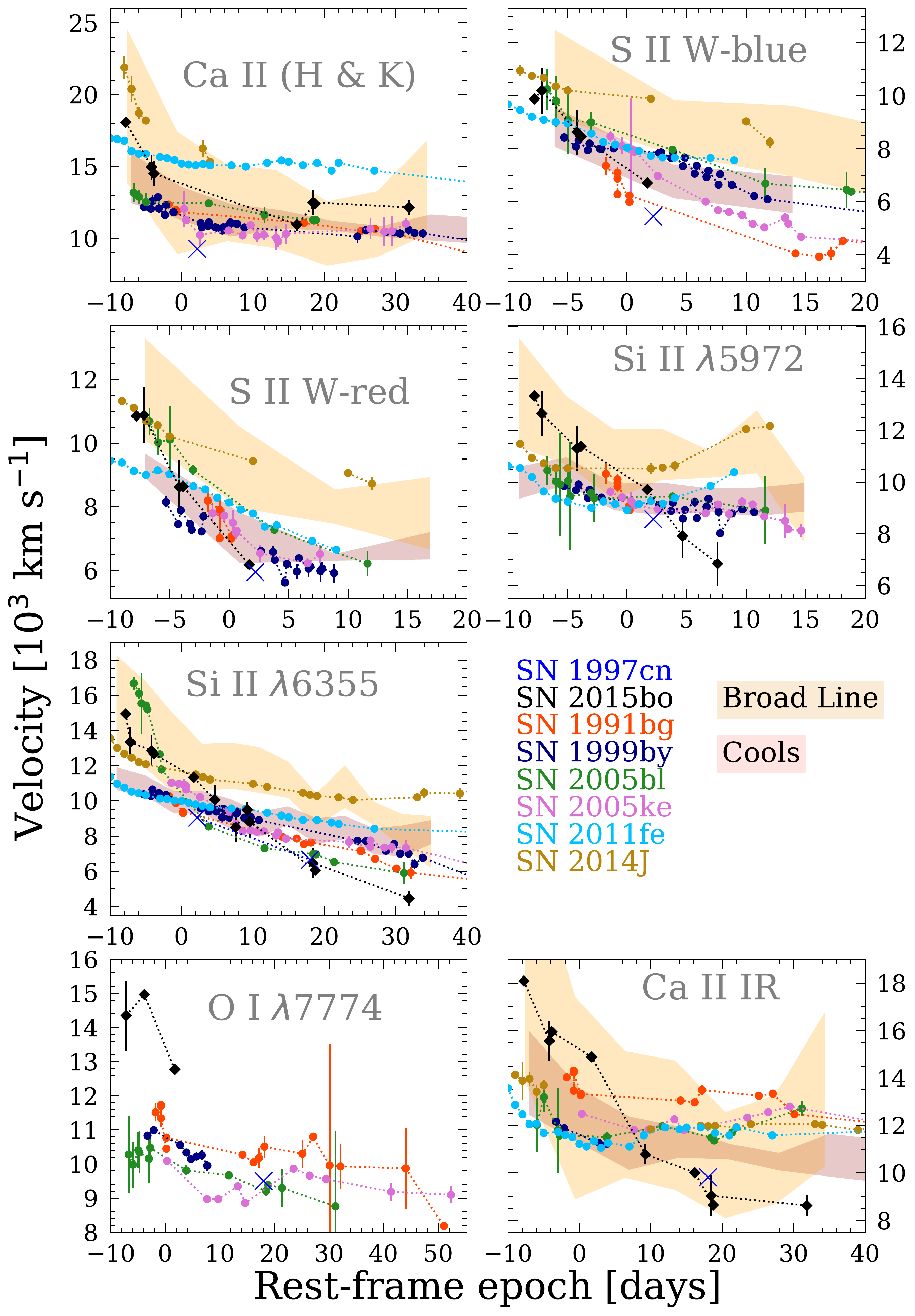}
\caption{Velocities of \cn\, (blue X), \bo\ (black, \nothing{diamonds}) and other \Ia\ (other colors, circles). The tan and salmon shaded areas are the 1$\sigma$ dispersion regions of the Broad Line and Cool evolutionary paths presented in \citet{Folatelli13} using CSP-I data. \label{fig:spec-vel}}
\end{figure*}

\subsubsection{Velocities}\label{sec:vel_vel_vel}
Expansion velocities for each strong spectral feature present in the spectra of \bo\ are shown in Figure \ref{fig:spec-vel} along with velocities from other \Ia\ for comparison. There are four sub-luminous SNe (SN~1991bg, SN~1999by, SN~2005bl, SN~2005ke) and two normal SNe (SN~2011fe and SN~2014J). Comparison velocities are taken from \citet{Galbany19}, and shaded regions represent the Cool and Broad Line regions from \citet{Folatelli13}. Uncertainties are derived from both the uncertainty in the measurement of the minimum of the feature and the resolution limit of the respective instrument used to observe each spectrum. These two uncertainties are added in quadrature to derive the final uncertainty. 

For the \CaII\ H\&K feature, \bo\ starts at $\sim$18,000~\kms\ $\sim$+8~d before \timeB\ near the normal SNe and decreases over time to reach $\sim$13,000~\kms\ at $\sim$+31~d after maximum to be among the sub-luminous SNe. This evolution does not mirror other sub-luminous \Ia\ in the sample, which have a flatter velocity evolution going from $\sim$13,000 \kms\ at early times to $\sim$11,000 \kms\ at later times. The evolution occupies the same space as the Cool region after $\sim-$5~d.

At early times, \SII\ W-blue feature's velocity is similar to SN~2005bl and SN~2014J. The velocity evolution follows the Cool and other 1991bg-like SNe. The evolution has an initial velocity of $\sim$11,000~\kms\ similar to SN~2005bl, however, this drops to $\sim$6,000 \kms\ which is similar to other sub-luminous \Ia. The \SII\ W-red feature shows analogous behavior; it starts near SN~2005bl and SN~2014J and declines over time to match the other sub-luminous Ia.

\bo\ has the fastest \SiII\ $\lambda$5972 feature of the sample at $\sim$13,500~\kms. This is in the Broad Line region. It then slows down until it is slower than all the other SNe. Unlike the \SII\ W-red feature, which reaches a homogenous value at $\sim-$5~d, the \SiII\ $\lambda$ 5972 feature continues to decline until it is the lowest velocity SN in the sample.

At $\sim-$8~d, \bo\ starts faster than all other sub-luminous \Ia\ \SiII\ $\lambda$6355 velocities with the exception of SN~2005bl at $\sim$15,000~\kms. After declining, the velocities are consistent with other fast-declining \Ia\ and are in the Cool region. At $\sim$+31~d, it is slower than any other member of the comparison sample with a velocity of $\sim$5,000\kms.

The two features of \SiII\ $\lambda$5972 and \SiII\ $\lambda$6355 exhibit velocities that diverge downwards from the Cool pathway of \citet{Folatelli13} between $\sim$0~d to $\sim$+5~d and $\sim$+10~d to $\sim$+20~d past maximum, respectively. These lower velocities also suggest that the base of the Silicon is deeper in \bo\ in all other \Ia\ in the sample. 

While noticeable residuals in the corrections for the telluric features are not seen in the epochs with calculated \OI\ $\lambda$7774 velocities, they still might exert influence on the calculations. This potential influence can be seen in significant difference between \bo\ and the rest of the comparison sample. However, if these measurements are correct and do not suffer from telluric absoption, then \bo\ has the fastest velocities of the sample. 

Finally, \bo's \CaII\ IR velocity is 18,000 \kms\ at $\sim-$10~d, the highest velocity in the entire sample. However, it declines rapidly from $\sim$0~d to $\sim$+10~d -- \bo\ transitions from being the fastest to the slowest in that time period. After $\sim$+10~d, \bo\ declines only slightly over the next twenty days to $\sim$8,500~\kms\ at +31~d after max.

Overall, the velocities are unique in their steep declines. \bo\ starts on the higher end of the \bg\ and then finishes on the lower end of the \bg\ distribution in nearly every panel of Figure \ref{fig:spec-vel}. Additionally, \bo\ primarily resides outside the Cool region. This behavior is unique to \bo. Along with the different $V-r$ color curve, this behavior gives evidence that \bo\ is different than other 1991bg-like objects.

\nothing{As is seen in the S and Si velocity evolution,   \bo\ has an extended layer of intermediate mass elements (IME). It is extended to both the highest and lowest velocity compared to SN of the same sub-type. The velocity of the IME region may imply that the burning near the surface was efficient leaving a lack of unburnt material and producing high velocity IME lines. Conversely, it also implies that density of burning in the inner layers (i.e. between 4000-6000~\kms) was less than about 1$\times 10^7$g cm$^{-3}$, above this density is where burning to Fe-group elements occur. It is unclear which explosion or progenitor scenarios can produce such a distribution.
However, within the frame work of a delayed detonation model the inner region is determined by the transition density (i.e. the density at which the flame front changes from a deflagration to a detonation), and the outer layer is determined by the C-O ratio within the WD (see e.g., \citealt{Hoeflich17}). }


\subsubsection{Pseudo-Equivalent Widths}\label{sec:pEW_pEW_pEW}

\begin{figure*}
\plotone{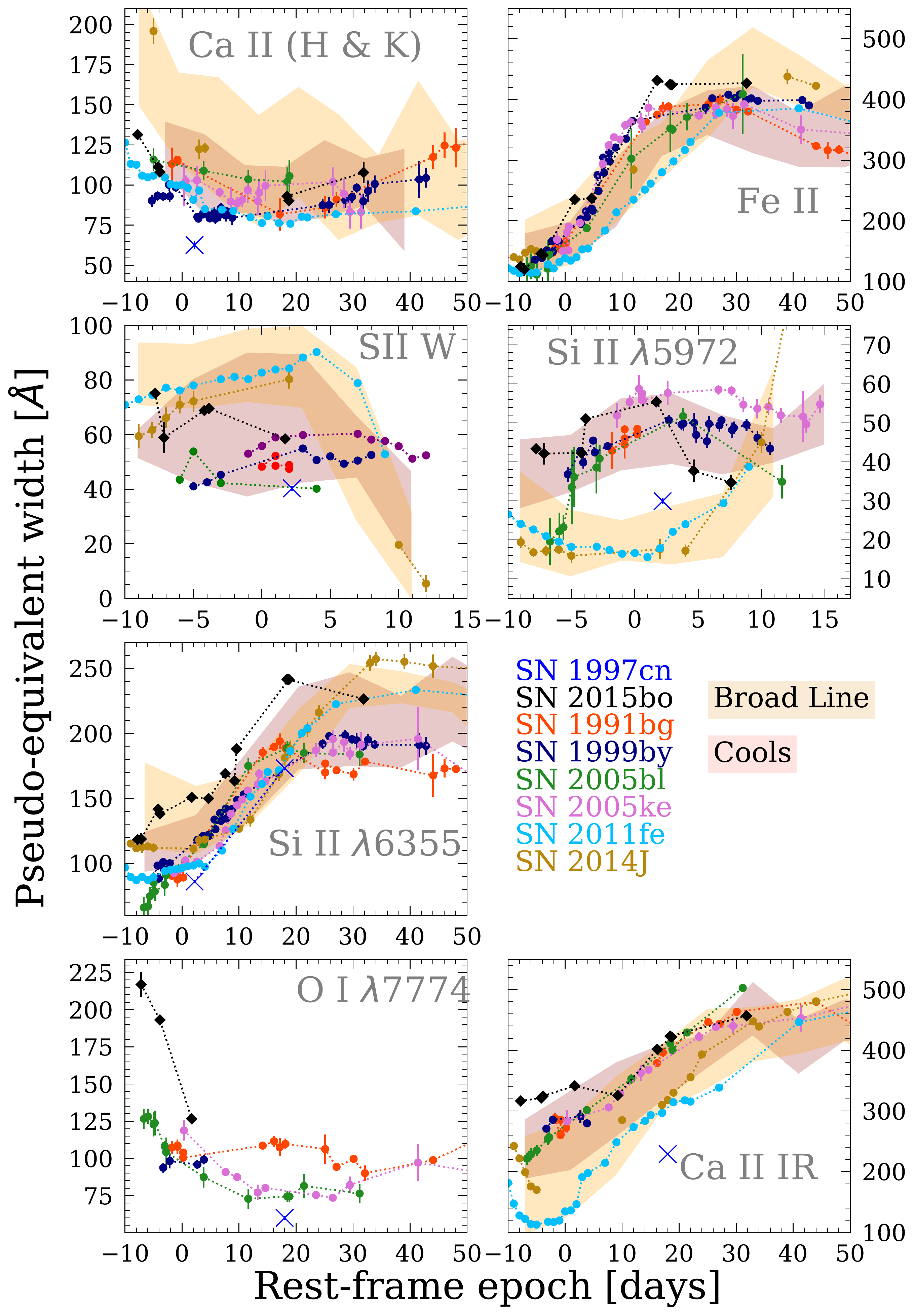}
\caption{Pseudo-equivalent widths of \cn\ (blue, X) \bo\ (black, \nothing{diamonds}) and other \Ia\ (other colors, circles). The tan and salmon shaded areas are the 1$\sigma$ dispersion regions of the Broad Line and Cool evolutionary paths presented in \citet{Folatelli13} using CSP-I data. \label{fig:spec-pew}}
\end{figure*}

The pEWs for \bo\ and selected comparison \Ia\ are plotted in Figure \ref{fig:spec-pew}. The pEW in \CaII\ H\&K for \bo\ is homogenous with the comparison sample even though \bo\ starts larger than any other Cool SN with a value of $\sim$130~\AA. The wavelength coverage of the spectra from $\sim$0~d to $\sim$+16~d did not fully capture the feature, so it is uncertain what value \bo\ would decline to, if at all. At $\sim$+31~d after maximum, \bo\ has the largest value at $\sim$115~\AA.

In \FeII, \bo\ is uniform with the rest of the sample, however, \bo\ exhibits a higher maximum value of $\sim$425~\AA\ at $\sim$+15~d.

The \SII~W pEW deviates from the rest of the sample; it starts near SN~2011fe at $\sim$75~\AA\ but drops below SN~2014J to $\sim$60~\AA\ at $\sim-$5~d. This is followed by a slight increase to $\sim$70~\AA\ five days prior to $B$-band maximum. At $\sim$+2~d after maximum, \bo\ declines down to $\sim$60~\AA. This evolution is not seen in any of the comparison SNe -- normal or sub-luminous. \bo\ inhabits the Cool region.

In \SiII\ $\lambda$ 5972, sub-luminous SNe have a concave down morphology and normal SNe have a concave up morphology. This is an effect of temperature and the heating of the photopshere. Since \bo\ is sub-luminous, it has a concave down form; it starts at $\sim$45~\AA, which is greater than the rest of the comparison collection, and then increases to a peak of $\sim$55~\AA\ at $\sim$+3~d after maximum. From there, \bo\ declines slightly to $\sim$35~\AA. The decline for \bo\ at $\sim$+5~d after maximum is earlier than the other sub-luminous SNe, but this time, perhaps coincidentally, corresponds to the rise of SN~2011fe. \bo\ occupies a similar area as the Broad Line region.

\bo\ has the same evolution curve as the rest of the sample in \SiII\ $\lambda$ 6355, however, its numerical values are marginally higher. \bo's maximum m$_B$ time at $\sim$+20~d after maximum is consistent with the peak of SN~1991bg and SN~2005bl, however, both SN~1999by and SN~2005ke peak later by $\sim$10~d. The largest pEW value of $\sim$235~\AA\ is larger than the four \Ia\ just mentioned -- they peak at $\sim$180-195~\AA. The only SN in the comparison sample that has a higher pEW is SN~2014J with a value of $\sim$255~\AA. This value occurs later than \bo's at $\sim$+35~d. Beyond $\sim$+25~d, we expect that the line is blended with \FeII\ and not just solely \SiII\ emission.  

In \OI\ $\lambda$ 7774, \bo\ starts well above the rest of the sample \Ia\ at almost 220~\AA\  $\sim-$8~d, but it returns to a more consistent value of $\sim$130~\AA\ at \timeB. The pEW in \OI\ may suffer from telluric absorption even though only epochs without obvious telluric contamination were selected. 

Finally, \bo\ starts marginally higher at $\sim$310~\AA\ at $\sim-$8~d in \CaII\ IR, but quickly integrates itself into the mean path of the sample $\sim$+10~d after maximum at $\sim$310~\AA\ and rises to reach $\sim$460~\AA\ at $\sim$+31~d after maximum.

\subsection{Near Infrared Spectrum}\label{sec:NIR}
\begin{figure}
\plotone{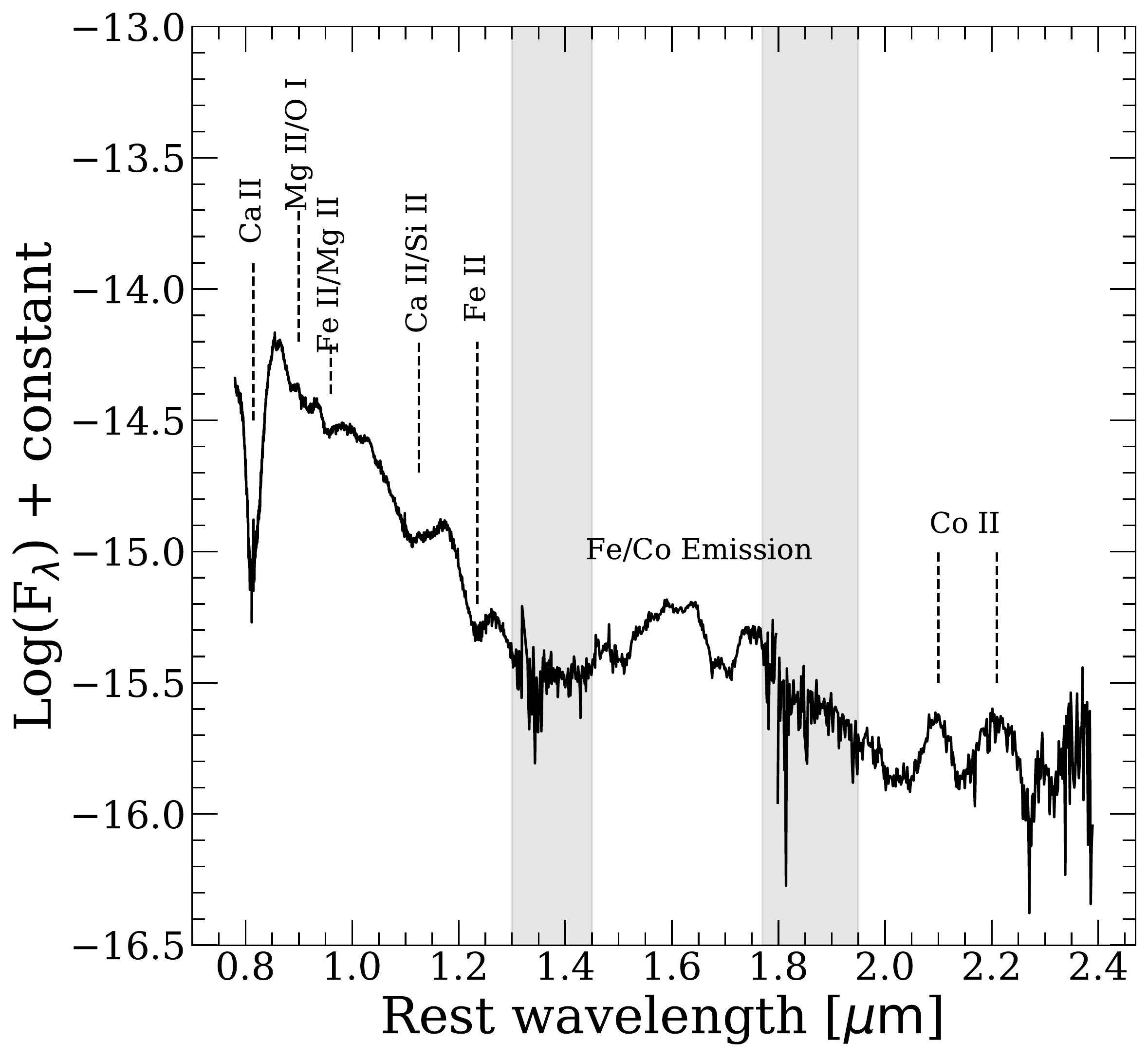}
\caption{NIR spectrum of \bo\ at +11.5~d. Telluric regions are denoted by grey bands. Spectral features are labeled on the plot and described in the text. \label{fig:spec-NIR}}
\end{figure}

The presented NIR spectrum was observed at +11.5~d and is shown in Figure \ref{fig:spec-NIR}. \nothing {We identify lines of both intermediate mass and Fe-group elements at this phase. Unlike the optical,  the NIR spectra reveal different layers of the ejecta at a given epoch \citep{Hoeflich02}.}
Specifically, we identify \CaII~$\lambda\lambda$8662, 8538, \MgII~$\lambda\lambda$9227,~10927, \SiII~$\lambda$11737, \CaII~$\lambda\lambda$11839,~11950, and \FeII\ features, as well as significant Fe and Co emission between the two telluric bands in the $H$-band break. Finally, strong \CoII\ emission is present in the 2~$\mu$m to 2.4~$\mu$m region. Eight \CoII\ lines dominate this region: \CoII\ $\lambda \lambda$20913, 21211, 21351, 21509, 22209, 22482, 23619, 24603 \citep{Gall12}.

\begin{figure}
\plotone{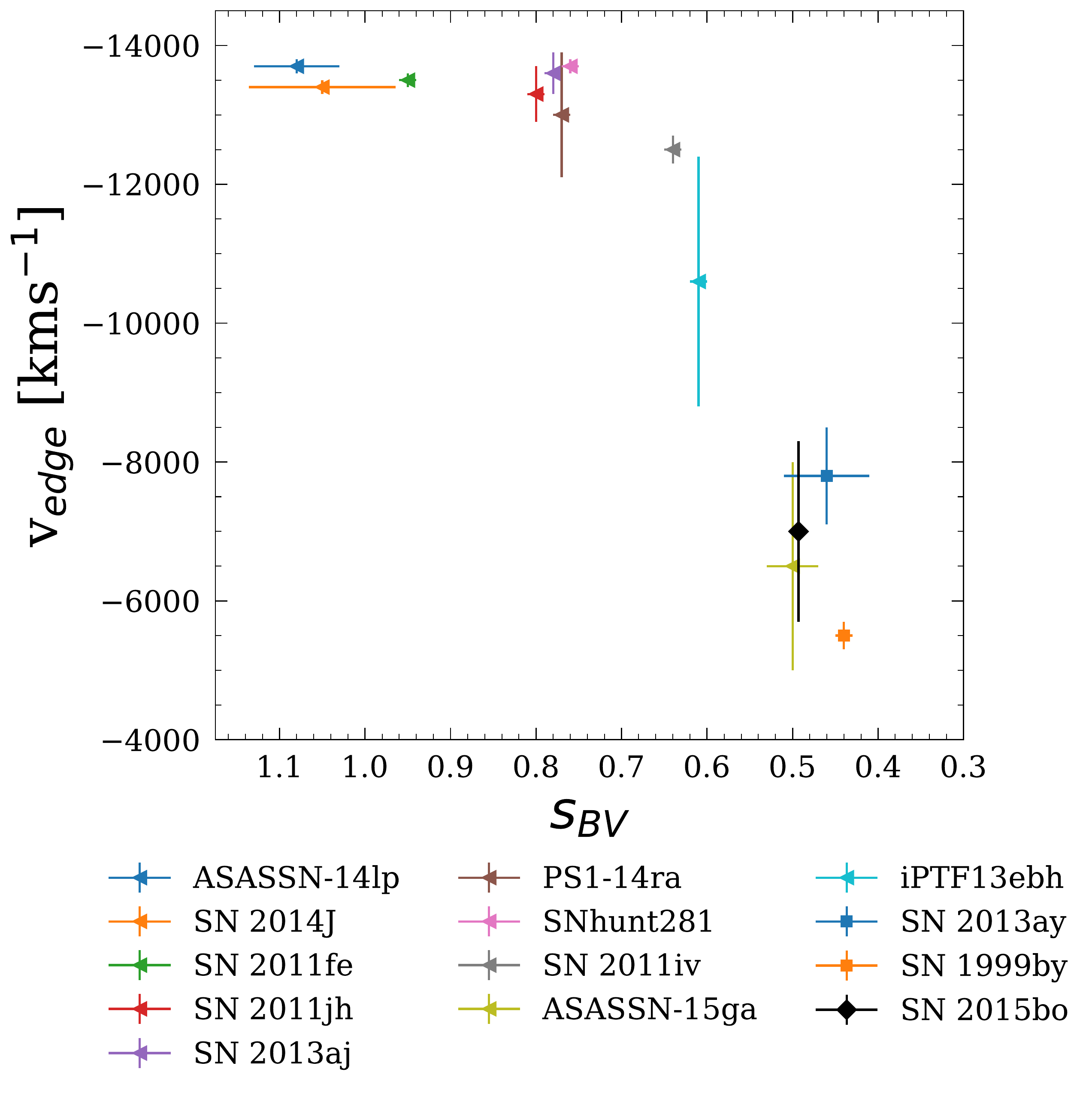}
\caption{\ved\ versus \sBV\ plot using data from \citet{Ashall19b}. $v_{edge}$ is the velocity of the bluest \CoII\ emission line in the $H$-band break. \nothing{\bo\ is plotted as the black diamond in the bottom right.} \label{fig:spec-vedge}}
\end{figure}

There is an area of strong \FeII, \CoII, and \NiII\ emission known as the $H$-band break that originates from allowed transitions above the photosphere \citep{Wheeler98}. \citet{Hsiao13} found a correlation between the SN stretch parameter and the feature strength of the $H$-band break and that the $H$-band break strength should be near its maximum value at the phase of our observed NIR spectrum. \citet{Ashall19a, Ashall19b} showed that this feature can be used to determine the location of the outermost \Nifs\ in \Ia\ using the $v_{edge}$ parameter. The $v_{edge}$ parameter is the velocity of the bluest edge of the $H$-band break \CoII\ emission line -- see \citet{Ashall19a} for further details. $v_{edge}$ is plotted against \sBV\ in Figure \ref{fig:spec-vedge} using data from \citet{Ashall19a}. 

\bo's $v_{edge}$ is $-7000 \pm 1300$ \kms\ \citep{Ashall19b}. This shows that the majority of the \Nifs\ mass is located in the center of the ejecta. This also implies a low \Nifs\ mass as found in §\ref{sec:bolo}.

\section{Distance to NGC 5490 \& 1991bg-LIKE SNE as Distance Indicators}\label{sec:dist}

\bo\ has a sibling (a SN in the same host galaxy): \cn\ \citep{Li97,tur98,jha06}. Interestingly, \bo\ and \cn\ are also both fast decliners, so not only are they siblings, \nothing{they are also of the same spectral sub-type (nominally called ``twins")}.  Since fast decliners are located preferentially in early-type galaxies \citep{Hamuy96}, it is unsurprising that \nothing{an early-type galaxy is the host of two fast-declining SNe as opposed to a different galaxy type where fast-declining SNe Ia are even less common}. This provides an opportunity to evaluate the accuracy and consistency of these quasi-rare sub-types as distance indicators. To date, this is the only known instance of two fast-declining SNe exploding in the same host galaxy. In this section, distance \nothing{moduli} are derived using various methods and checked for consistency. 

We calculate the distance modulus to the host (NGC~5490) with three independent methods: i) the $z_{cmb}$ distance derived method assuming a cosmology of $H_{0}$=73\,km\,s$^{-1}$\,Mpc$^{-1}$, $\Omega_{m}$=0.27 and $\Omega_{\Lambda}$=0.73, ii) the independent distance derived using the photometry of all available bands of \bo, and iii) the independent distance derived using the photometry of all available bands of \cn. All of the derived distances, as well as the distance computed using surface-brightness fluctuations \citep{Jensen21}, can be found in Table \ref{tab:dist-mods}. 

\begin{deluxetable}{ccc}
\tablenum{4}
\label{tab:dist-mods}
\tablecaption{Distance \nothing{moduli} derived using four independent methods: $z_{cmb}$ cosmology, surface-brightness fluctuations, SN~2015bo, and SN~1997cn.}
\tablewidth{0pt}
\tablehead{
\colhead{Object} & \colhead{$\mu$} & \colhead{$\sigma_{\mu}$} \\ 
    & [mag] & [mag]}
\startdata
NGC 5490\tablenotemark{a} & 34.21 & 0.15\\
NGC 5490\tablenotemark{b} &  34.27 & 0.08 \\
SN~2015bo & 34.33 & 0.01 (stat) 0.11 (sys) \\
SN~1997cn & 34.34 & 0.04 (stat) 0.12 (sys)
\enddata
\tablenotemark{a}{Corrected to the CMB rest frame and calculated using $H_{0}$=73\,km\,s$^{-1}$\,Mpc$^{-1}$, $\Omega_{m}$=0.27 and $\Omega_{\Lambda}$=0.73, which is used throughout this work.} \\
\tablenotemark{b}{Distance derived from surface brightness fluctuation distance from \citet{Jensen21}}
\end{deluxetable}

The distance modulus derived using the $z_{cmb}$ method is $\mu=34.21 \pm\ 0.15$~mag. As discussed in §\ref{sec:lc}, the distance modulus to \bo\ was determined using the  EBV2 model and a 1991bg-like K-correction in \snpy\ \nothing{with the prior that $E(B-V)_{host}=0$ in the fit}. Photometry for \bo\ is in the CSP natural system. \snpy\ has an uncertainty budget described in \citet{Burns11}. Of the four categories, only the uncertainty from the stretch parameter applies to \bo. To compute the systematic uncertainty for \sBV, the dispersion in the fit between \sBV\ and \DmB\ is considered. Because these are independent measures of the decline rate, the source of the dispersion in this relation comes from both random noise (due to measurement error) and any systematic errors that \sBV\ and \DmB\ have in measuring the “true” decline rate. Therefore the dispersion is an upper limit on the systematic error in \sBV\ (or \DmB). This value is $\sigma s_{BV}=0.03$, which turns into an uncertainty in distance modulus of $\pm0.11$~mag. Thus, we calculate a distance modulus of $\mu$~=~\nothing{34.33}~$\pm$~0.01~(stat)~$\pm$~0.11~(sys)~mag. 

\begin{figure}
\plotone{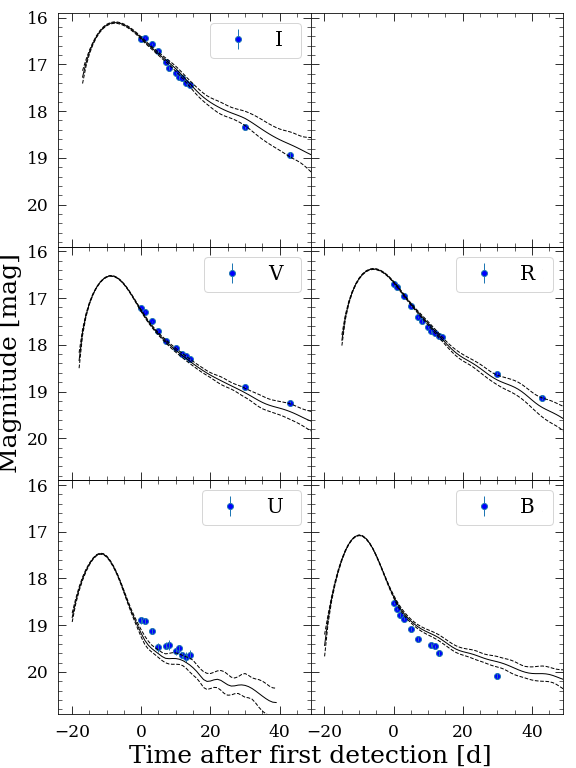}
\caption{SNooPy fits for SN~1997cn. Data from \citet{jha06}. The fits are constrained by using the \timeB\ from SNID. \nothing{Host-galaxy extinction is a free parameter for these fits}. \label{fig:snpy-97cn-jha}}
\end{figure}

The standard system filters are used when fitting the photometry of \cn\ with \snpy\ because the data for \cn\ is in the standard system. Since there is not enough photometric data to accurately determine \timeB, the best spectrum for \cn\ was used as input in SNID to determine the time of maximum using the spectrum. The best SNID match was to \nothing{the 1991bg-like SN~1998de \citep{Modjaz01}} at +2.7~d old.  The time from this match (\timeB = 50587.8 MJD) was then given to \snpy\ to use as the time of $B$-band maximum. Without constraining \timeB, SNooPy struggles to reasonably fit the light curves. This is because SNooPy seeks to fit the $I$-band with two maxima like normal \Ia, however, \cn\ does not have a second maximum. This causes the fit to be poor, \nothing{so the fit is given the \timeB\ from SNID as a prior. We also choose to leave host-galaxy extinction as a free parameter for \cn.} Apart from these differences, the fitting methods used to derive the distance modulus of \cn\ are the same as the ones used to derive the distance modulus of \bo. Only the data from \citet{jha06} was used because it was obtained on a single telescope and published in the standard system. \citet{tur98} use data from multiple telescopes and it is unclear what system the data is in. Fits for \cn\ are presented in Figure \ref{fig:snpy-97cn-jha}.

To calculate the total \snpy\ systematic uncertainty in \cn, we must consider the systematic uncertainty from reddening in addition to the intrinsic spread in \sBV. The reddening uncertainty was determined via two methods. The first method took the systematic uncertainty as the difference between distance modulus when calculated with $E(B-V)_{host}$ as a free parameter and held fixed to a value of 0. This gave a systematic uncertainty in distance modulus of $\pm$0.01~mag for \cn. The second method was to use the intrinsic dispersion in the B-V intrinsic color. The dispersion is $\pm$0.06, so one could also simply adopt that as a systematic. This gives an uncertainty of $\pm$0.04~mag for \cn. The method with the larger value was chosen to be used in the final analysis and was added in quadrature with the \sBV\ systematic uncertainty to arrive at a final distance modulus of $\mu$~=~\nothing{34.34}~$\pm$~0.04~(stat)~$\pm$~0.12~(sys)~mag. 

Adding the statistical and systematic errors in quadrature for \cn\ and \bo\ gives values of $\mu_{97cn} = 34.34 \pm 0.13$~mag and $\mu_{15bo} = 34.33 \pm 0.11$~mag. The difference in distance modulus is $\Delta\mu = 0.01 \pm 0.17$~mag, again with the errors added in quadrature. The distances differ by less than 0.06-$\sigma$.

\begin{figure}
\plotone{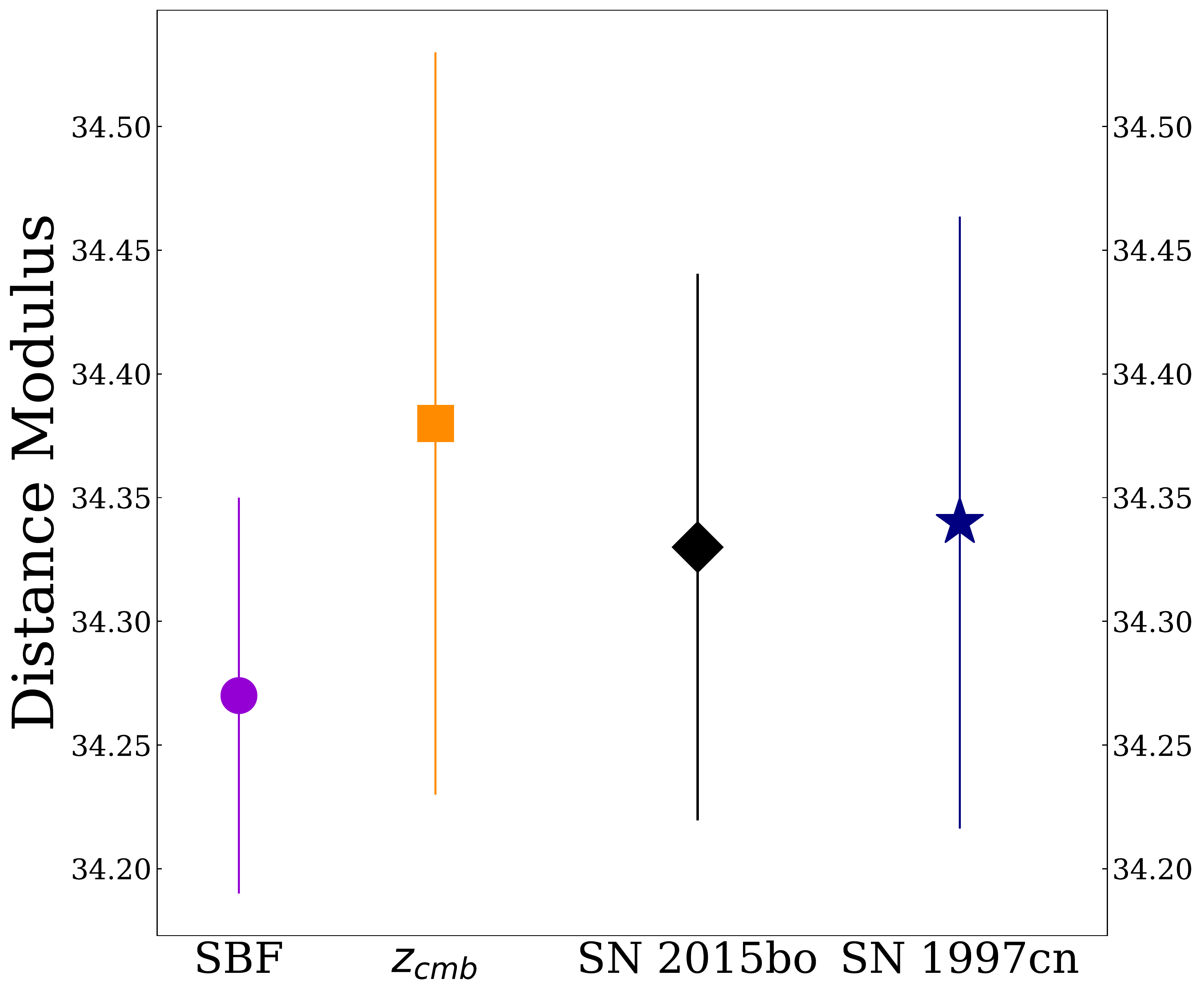}
\caption{Graphical comparison of distance \nothing{moduli} derived from surface brightness fluctuation (SBF; taken from \citealt{Jensen21}), $z_{cmb}$-corrected cosmology, \bo, and \cn. All values are consistent with each other. \label{fig:dm-comp}}
\end{figure}

Figure \ref{fig:dm-comp} provides a visual representation of the distance \nothing{moduli} where the consistency is clearly seen. Additionally, distance \nothing{moduli} from both redshift-corrected cosmology and surface-brightness fluctuations from \citet{Jensen21}. Note that there is most likely further uncertainty due to fits for \cn\ being poorly constrained from the lack of data prior to maximum in all bands. This will add in an extra uncertainty in the value for \cn. Therefore, the presented values are lower bounds for the uncertainty. Considering this, it appears that fast decliners might be equally standardizable as normal \Ia\ and could offer the same robust functionality. The result from \citet{Burns18} supports this hypothesis: they found SN~2006mr to be consistent with the three other \Ia\ in Fornax A \citet{Stritzinger10}.

 However, there have been inconsistencies in the past with using twins and siblings when deriving distances. \citet{Foley20} found a significant ($\delta \mathrm{M}_V = 0.335 \pm 0.069$~mag) difference between the peak $V$-band magnitudes of 2011by and 2011fe, but these SNe were hosted by different galaxies -- NGC 3972 and M101, respectively -- they may be influenced by peculiar velocities. \citet{Gall18} found that two transitional supernovae (SN~2007on and SN~2011iv) that exploded in the same host galaxy produce a statistical difference in  calculated distance \nothing{moduli} in both $B$- and $H$-bands. They found the two objects' distance \nothing{moduli} differed by $\sim$14\% and $\sim$9\% in the $B$- and $H$-bands, respectively. On the other hand, \bo\ and \cn\ are in the same host galaxy, have similar spectra, and have consistent distances. This tension can be resolved with further twin studies of fast decliners.
 
\begin{figure}
\plotone{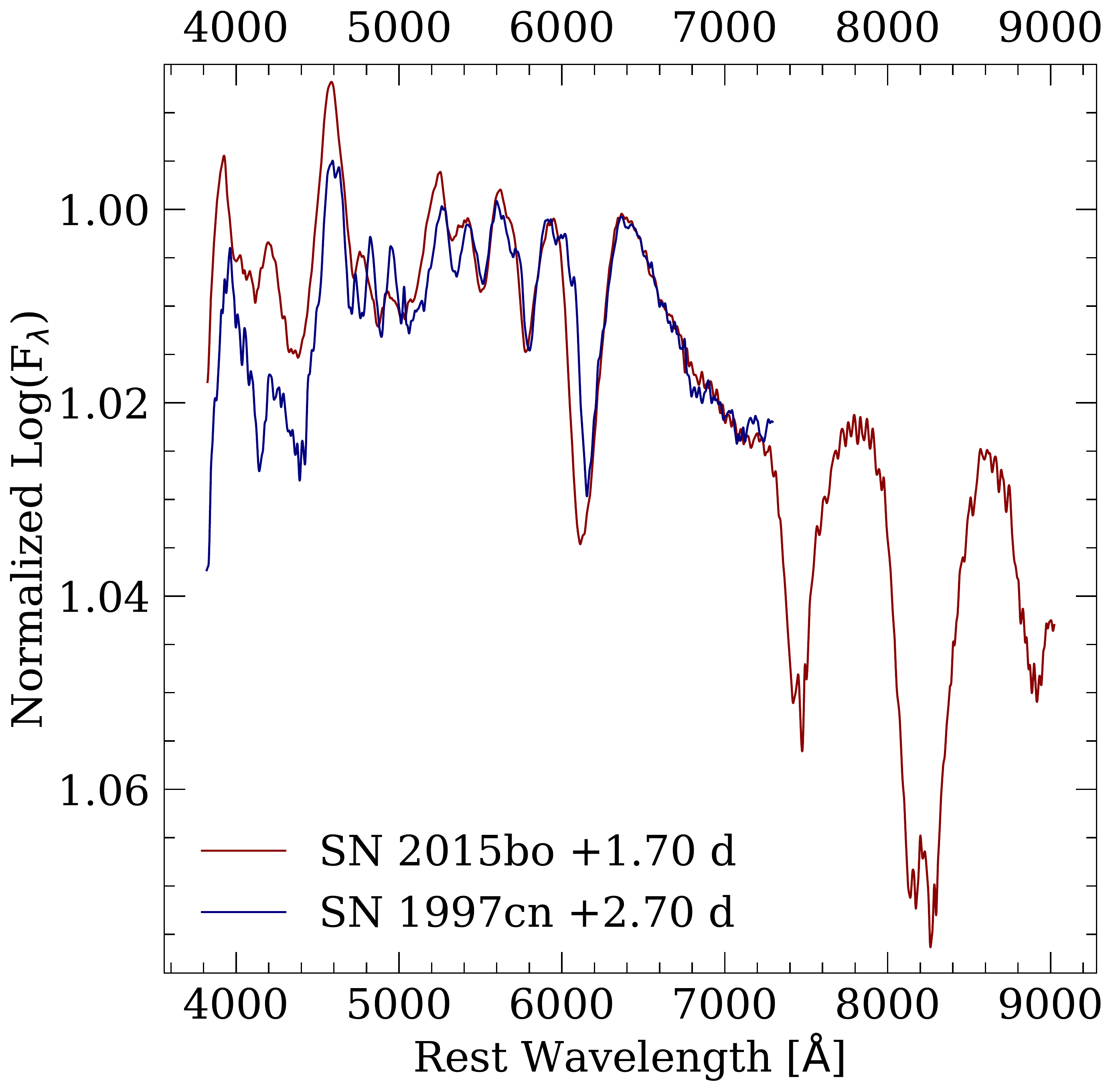}
\caption{Comparison of the spectra of \cn\ (navy) and \bo\ (dark red). The spectra have been color corrected to match the photometric measurements, put through a Gaussian 3-$\sigma$ smoothing function, and were normalized to the blue edge of the \SiII\ $\lambda$ 6355 feature.  \label{fig:bg-spec-comp}}
\end{figure}

 The maximum light spectral similarity of \cn\ and \bo\ can be seen in Figure \ref{fig:bg-spec-comp}. The `twinness' parameter described by \citet{Fakhouri15} was calculated to be 1.04, which puts it in the median range of twinness. This parameter is akin to a reduced $\chi^2$ parameter, and therefore it is sensitive to estimated errors. This estimated error dominates the spectra possibly due to the data being taken on different telescopes and instruments, and thus the twinness parameter may not be the best indicator of how alike two spectra or \Ia\ are unless the quality of the input data is similar.

\section{Conclusion} \label{sec:conclusion}

Photometric $uBVgriYHJ$ data of \bo\ from $-$2.9~d to +103.8~d and spectroscopic data from $-$7.8~d to 31.8~d from maximum light are presented. \bo\ is a  fast-declining, under-luminous SN Ia that exploded in NGC 5490. \bo\ has an absolute magnitude \nothing{$M_B = -17.67$} $\pm$ \nothing{0.15}~mag. It has \DmB = 1.91$\pm$0.01~(mag) and \sBV = 0.48 $\pm$ 0.01. 

Photometrically, \bo\ is sub-luminous and fast-declining, and it occupies the same parameter space as other under-luminous, fast decliners in the luminosity-width relation. This is confirmed in the \sBV\ versus $t^{i-B}_{max}$ diagram where \bo\ is positioned among \bg. The light-curves evolve quickly, and \bo\ has unique morphologies in the $V-r$ and $r-i$ color-curves. In the $V-r$ color-curve, it is the bluest SNe in the plotted sample even with extinction corrections applied. In the $r-i$ color-curve, it displays a redward climb at $\sim$+40~d. This may be a way to identify peculiar sub-luminous SNe in future surveys. \bo\ has a peak bolometric luminosity of $L^{peak}_{bolo}=0.33 \pm 0.03 \times 10^{43}$~\ergs, which corresponds to a nickel mass of $0.17 \pm 0.05 {M_\odot}$. This value may change slightly if the full Arnett's rule and a rise time were used. 

Spectroscopically, \bo\ shows the standard features typical for a fast decliner. In the \citet{Branch06} diagram, \bo\ resides among other Cool \Ia. \bo's velocities range from $\sim$18,000\kms to $\sim$4,000\kms\ and decrease over time. 

Importantly, \bo\ exploded in the same host galaxy as \cn, another fast decliner. Distances to both \bo\ and \cn\ were calculated and are consistent with each other at the 0.06-$\sigma$ level.  We suggest that fast decliners are consistent and standardizable, and hence they should \emph{not} be omitted from future cosmological work. A future study of two or more fast-declining siblings with two or more well sampled objects will offer a chance to confirm this result. 

\appendix
\restartappendixnumbering
\renewcommand{\thefigure}{A\arabic{figure}}
\renewcommand{\theHfigure}{A\arabic{figure}}
\setcounter{figure}{0}
\section{\cn\ Light Curves}\label{sec:A}
\begin{figure}
\plotone{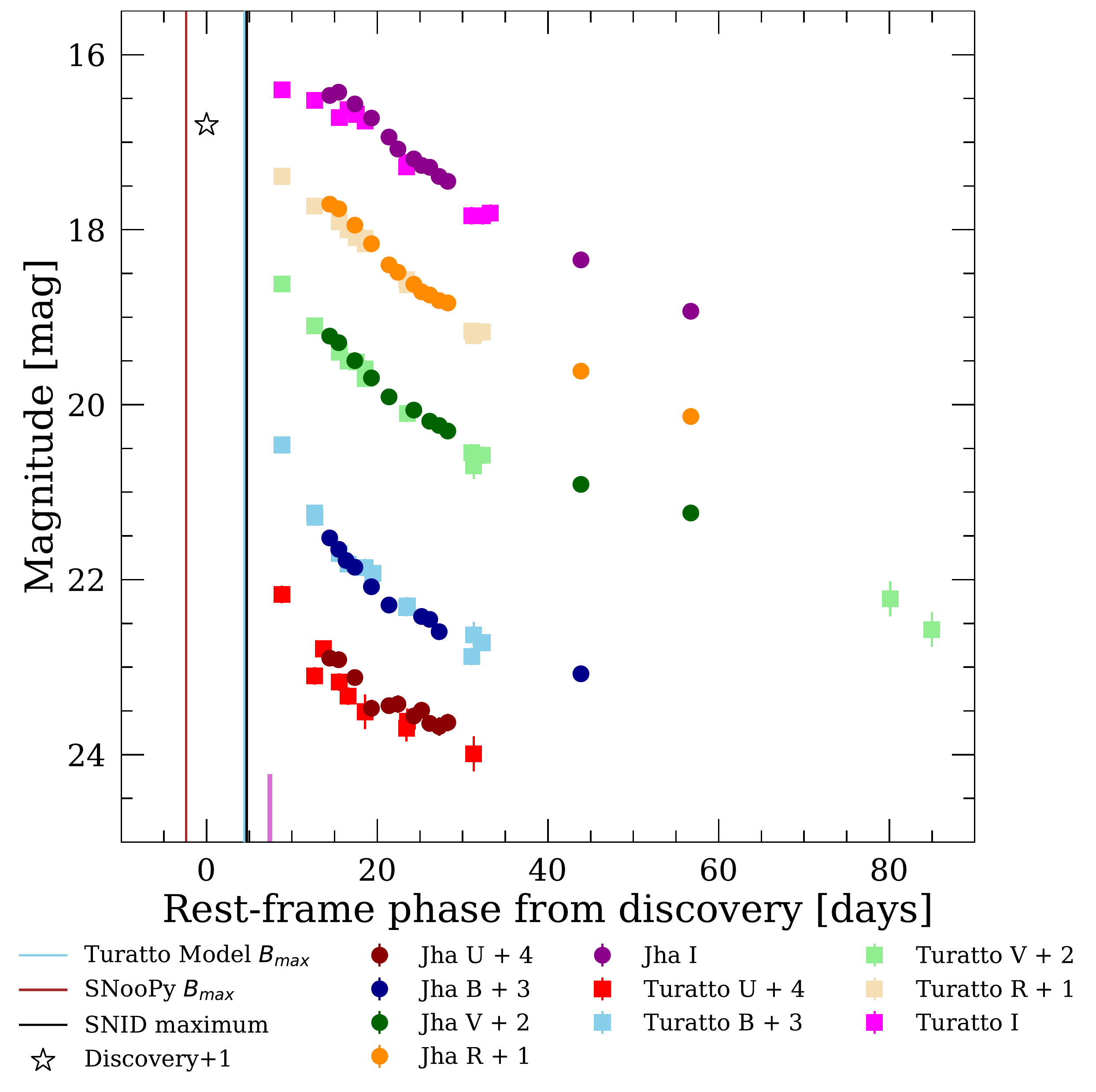}
\caption{Light curves of SN~1997cn. The discovery datum point is from \citealt{Li97} and is denoted by a hollow black star, lightly filled, square data points are from \citet{tur98}, and darkly filled, circle data points are from \citet{jha06}. The discovery point from \citet{Li97} uses an unfiltered CCD exposure. This point is plotted as if it was in the $R$-band. The magenta line represents the date of the earliest spectrum for \cn. \label{fig:lc_97cn}}
\end{figure}

Light curves for SN~1997cn are presented in Figure \ref{fig:lc_97cn}, as well as \timeB\ as derived from \snpy, SNID spectrum analysis, and the spectral modelling performed by \citet{Turatto98}. \citet{tur98} estimate $B$-band maximum occurred on 1997 May 19 (2450587.50 JD) which is +18~d after the inferred explosion. The time derived by SNooPy using data from \citet{jha06} is 2450580.72~$\pm$~0.70~JD. The difference is $\Delta$\timeB$=6.78$~d. The \snpy\ \timeB\ is before the discovery point, however, the discovery point could be close to maximum. Given that the data was taken without a filter, it most likely resembles $V$- or $R$-band photometry. The SNID \timeB\ is 2450588.3~JD. This time yields the best SNooPy fits of the light curves. SNooPy differs from these dates because there is no well-defined maximum in the light curves for \cn. The exact date of maximum is uncertain. This comes from the fact that comprehensive, all-sky surveys for SNe were not available when SN~1997cn exploded.

\renewcommand{\thefigure}{B\arabic{figure}}
\renewcommand{\theHfigure}{B\arabic{figure}}
\renewcommand{\thetable}{B\arabic{table}}
\renewcommand{\theHtable}{B\arabic{table}}
\section{Photometric Data}\label{sec:data-tabs}

\startlongtable
\begin{deluxetable*}{ccccccc} 
\tablenum{B1}
\tablecaption{Log of optical photometric data.}
\label{tab:appendix-phot-data-optical}
\tablewidth{0pt}
\tablehead{
\colhead{day}&\colhead{$u$}&\colhead{$B$}&\colhead{$V$}&\colhead{$g$}&\colhead{$r$}&\colhead{$i$} \\
\colhead{[MJD]}&\colhead{[mag]}&\colhead{[mag]}&\colhead{[mag]}&\colhead{[mag]}&\colhead{[mag]}&\colhead{[mag]}}
\startdata
57069.4&18.51(0.03)&17.49(0.01)&17.30(0.01)&17.35(0.01)&17.21(0.01)&17.49(0.01) \\
57075.4&18.16(0.02)&16.84(0.01)&16.52(0.01)&16.63(0.01)&16.37(0.01)&16.66(0.01) \\
57076.3&18.24(0.02)&16.86(0.01)&16.47(0.01)&16.63(0.01)&16.32(0.01)&16.65(0.01) \\
57079.4&18.55(0.03)&17.08(0.01)&16.48(0.01)&16.76(0.01)&16.28(0.01)&16.64(0.01) \\
57080.3&18.66(0.03)&17.22(0.01)&16.51(0.01)&16.86(0.01)&16.30(0.01)&16.65(0.01) \\
57082.3&$\cdots$&17.55(0.01)&16.66(0.01)&17.13(0.01)&$\cdots$&$\cdots$ \\
57082.4&19.07(0.03)&$\cdots$&$\cdots$&$\cdots$&16.37(0.01)&16.72(0.01) \\
57083.3&19.22(0.04)&17.71(0.01)&16.74(0.01)&17.28(0.01)&16.44(0.01)&16.75(0.01) \\
57084.3&19.41(0.05)&17.90(0.02)&16.84(0.01)&17.46(0.02)&16.50(0.01)&16.79(0.02) \\
57085.3&19.62(0.06)&18.05(0.02)&16.96(0.02)&17.59(0.02)&16.56(0.01)&16.78(0.01) \\
57086.3&$\cdots$&$\cdots$&17.10(0.02)&17.76(0.02)&16.65(0.01)&16.83(0.01) \\
57086.4&$\cdots$&18.15(0.02)&$\cdots$&$\cdots$&$\cdots$&$\cdots$ \\
57087.3&$\cdots$&18.32(0.03)&17.16(0.01)&17.93(0.02)&16.71(0.01)&16.83(0.01) \\
57087.4&19.76(0.15)&$\cdots$&$\cdots$&$\cdots$&$\cdots$&$\cdots$ \\
57088.3&$\cdots$&18.45(0.02)&17.27(0.02)&18.02(0.02)&16.77(0.01)&16.86(0.02) \\
57089.3&$\cdots$&18.59(0.03)&17.37(0.02)&18.18(0.02)&16.85(0.01)&16.88(0.01) \\
57090.4&$\cdots$&18.70(0.03)&17.45(0.02)&18.27(0.02)&16.92(0.01)&16.93(0.01) \\
57091.3&$\cdots$&18.79(0.02)&17.56(0.01)&18.36(0.02)&16.99(0.01)&16.97(0.01) \\
57092.4&$\cdots$&18.83(0.03)&17.65(0.02)&18.42(0.02)&17.06(0.01)&17.04(0.01) \\
57093.3&$\cdots$&19.00(0.02)&17.70(0.01)&18.54(0.02)&17.16(0.01)&17.10(0.01) \\
57094.3&$\cdots$&19.01(0.02)&17.82(0.02)&$\cdots$&$\cdots$&$\cdots$ \\
57094.4&$\cdots$&$\cdots$&$\cdots$&18.62(0.02)&17.23(0.01)&17.17(0.01) \\
57095.3&$\cdots$&19.08(0.02)&17.88(0.01)&18.68(0.02)&$\cdots$&$\cdots$ \\
57095.4&$\cdots$&$\cdots$&$\cdots$&$\cdots$&17.32(0.01)&17.24(0.01) \\
57096.3&$\cdots$&19.14(0.02)&17.97(0.02)&$\cdots$&$\cdots$&$\cdots$ \\
57096.4&$\cdots$&$\cdots$&$\cdots$&18.75(0.02)&17.42(0.01)&17.32(0.01) \\
57097.3&$\cdots$&$\cdots$&$\cdots$&18.83(0.02)&17.52(0.01)&$\cdots$ \\
57097.4&$\cdots$&19.24(0.02)&18.04(0.02)&$\cdots$&$\cdots$&17.44(0.01) \\
57098.3&$\cdots$&$\cdots$&$\cdots$&18.88(0.02)&17.62(0.01)&17.54(0.01) \\
57098.4&$\cdots$&19.25(0.02)&18.14(0.02)&$\cdots$&$\cdots$&$\cdots$ \\
57099.3&$\cdots$&19.32(0.02)&18.20(0.02)&18.89(0.02)&17.70(0.01)&17.62(0.01) \\
57100.3&$\cdots$&19.39(0.02)&18.27(0.02)&19.00(0.02)&17.79(0.01)&17.71(0.01) \\
57101.3&$\cdots$&19.46(0.02)&18.33(0.01)&19.04(0.02)&17.84(0.01)&17.79(0.01) \\
57102.3&$\cdots$&19.49(0.02)&18.38(0.02)&19.11(0.02)&17.89(0.01)&17.84(0.02) \\
57103.4&$\cdots$&$\cdots$&18.42(0.04)&19.13(0.04)&17.81(0.06)&$\cdots$ \\
57104.3&$\cdots$&$\cdots$&$\cdots$&$\cdots$&18.00(0.02)&$\cdots$ \\
57109.3&$\cdots$&$\cdots$&$\cdots$&$\cdots$&18.23(0.01)&$\cdots$ \\
57109.4&$\cdots$&19.67(0.02)&18.62(0.02)&19.27(0.02)&$\cdots$&18.24(0.02) \\
57110.2&$\cdots$&$\cdots$&$\cdots$&19.34(0.02)&18.29(0.01)&$\cdots$ \\
57110.3&$\cdots$&19.69(0.02)&18.68(0.02)&$\cdots$&$\cdots$&18.30(0.02) \\
57111.3&$\cdots$&19.76(0.02)&18.71(0.02)&19.36(0.02)&18.33(0.01)&18.39(0.02) \\
57112.3&$\cdots$&19.66(0.02)&18.73(0.02)&19.37(0.02)&18.38(0.01)&18.39(0.01) \\
57113.3&$\cdots$&19.74(0.05)&18.77(0.02)&19.37(0.04)&18.42(0.02)&18.47(0.02) \\
57117.2&$\cdots$&$\cdots$&$\cdots$&19.40(0.06)&18.58(0.03)&$\cdots$ \\
57117.3&$\cdots$&19.76(0.09)&18.91(0.05)&19.45(0.06)&$\cdots$&18.64(0.03) \\
57119.3&$\cdots$&19.76(0.07)&18.96(0.04)&$\cdots$&18.64(0.02)&18.73(0.03) \\
57121.2&$\cdots$&19.92(0.04)&19.00(0.02)&19.54(0.03)&18.72(0.02)&18.79(0.02) \\
57123.3&$\cdots$&20.00(0.04)&19.05(0.02)&19.55(0.02)&18.82(0.02)&18.88(0.02) \\
57126.2&$\cdots$&20.03(0.02)&19.17(0.02)&$\cdots$&$\cdots$&$\cdots$ \\
57126.3&$\cdots$&$\cdots$&$\cdots$&19.64(0.02)&18.98(0.02)&19.04(0.02) \\
57129.2&$\cdots$&20.05(0.02)&19.26(0.02)&19.67(0.02)&19.08(0.02)&19.16(0.02) \\
57131.3&$\cdots$&20.07(0.02)&19.32(0.02)&19.70(0.02)&19.15(0.02)&19.21(0.02) \\
57134.3&$\cdots$&20.10(0.03)&19.38(0.02)&19.77(0.02)&19.30(0.02)&19.30(0.03) \\
57136.3&$\cdots$&20.13(0.03)&19.42(0.02)&19.81(0.02)&19.35(0.02)&19.40(0.03) \\
57138.2&$\cdots$&20.16(0.03)&19.50(0.02)&19.84(0.02)&19.42(0.02)&19.45(0.02) \\
57153.2&$\cdots$&20.43(0.04)&19.85(0.03)&20.10(0.03)&20.05(0.03)&19.94(0.05) \\
57160.2&$\cdots$&20.61(0.04)&20.04(0.03)&20.18(0.03)&20.26(0.04)&20.15(0.07) \\
57180.1&$\cdots$&$\cdots$&$\cdots$&20.61(0.04)&20.98(0.07)&$\cdots$ \\
57181.2&$\cdots$&$\cdots$&$\cdots$&20.50(0.09)&$\cdots$&$\cdots$ \\
57182.1&$\cdots$&$\cdots$&20.38(0.09)&20.60(0.07)&$\cdots$&$\cdots$ \\
\enddata
\end{deluxetable*}

\startlongtable
\begin{deluxetable*}{cccc}
\tablenum{B2}
\tablecaption{Log of near-infrared photometric data.}
\label{tab:appendix-phot-data-NIR}
\tablewidth{0pt}
\tablehead{
\colhead{day}&\colhead{$J$}&\colhead{$H$}&\colhead{$Y$} \\
\colhead{[MJD]}&\colhead{[mag]}&\colhead{[mag]}&\colhead{[mag]}}
\startdata
57088.3&16.97(0.02)&16.51(0.02)&16.30(0.01) \\
57089.4&16.97(0.03)&16.53(0.02)&16.31(0.01) \\
57090.4&17.00(0.03)&16.59(0.02)&16.31(0.01) \\
57091.3&17.06(0.03)&16.57(0.02)&16.32(0.01) \\
57093.3&17.15(0.02)&16.72(0.02)&16.35(0.01) \\
\enddata
\end{deluxetable*}

\acknowledgements{
We thank the anonymous referee for their comments.  W.B.H. acknowledges support from Research Experience for Undergraduates program at the Institute for Astronomy, University of Hawaii-Manoa funded through NSF grant \#2050710. 
L.G. acknowledges financial support from the Spanish Ministry of Science, Innovation and Universities (MCIN) under the 2019 Ram\'on y Cajal program RYC2019-027683 and from the Spanish MCIN project HOSTFLOWS PID2020-115253GA-I00.
M. G. M., R. G. D., and S. M. T. were funded by the European Union's Horizon 2020 research and innovation programme under the Marie Sk\l{}odowska-Curie grant agreement No. 839090.
Based on observations collected at the European Southern Observatory under ESO programme 106.2104.
The work of the CSP-II has been generously supported by the National Science Foundation under grants AST-1008343, AST-1613426, AST-1613455, and AST-1613472. The CSP-II was also supported in part by the Danish Agency for Science and Technology and Innovation through a Sapere Aude Level 2 grant.
E.B. was partially supported by NASA Grant 80NSSC20K0538
J.D.L acknowledges support from a UK Research and Innovation Fellowship (MR/T020784/1).
C.R.B. acknowledges support from NSF grants AST-1008384, AST-1613426, AST-1613455, and AST-1613472.
}
M.S and S. H. are supported by grants from the VILLUM FONDEN (grant number 28021) and  the Independent Research Fund Denmark (IRFD; 8021-00170B).

\bibliography{sample631}{}
\bibliographystyle{aasjournal}

\end{document}